\documentclass[12pt, a4paper]{article}
\usepackage[utf8]{inputenc}
\usepackage[english]{babel}
\usepackage[a4paper,top=2cm,bottom=2.5cm,left=2.5cm,right=2.5cm,marginparwidth=1.75cm]{geometry}
\usepackage{amsmath,amsfonts}
\usepackage{amssymb}
\usepackage{caption,adjustbox}
\usepackage{subcaption}
\usepackage{graphicx}
\usepackage[colorlinks=true,linkcolor=NavyBlue,citecolor=NavyBlue,urlcolor=Blue]{hyperref}
\usepackage[usenames,dvipsnames]{xcolor}
\usepackage{physics,tensor,cancel}
\usepackage{tikz}
\usetikzlibrary{calc}
\usepackage[affil-sl]{authblk}

\usepackage{comment}
\usepackage{csquotes}
\usepackage{hyperref}
\usepackage[normalem]{ulem}
\usepackage{fancyhdr}
\usepackage{academicons}
\usepackage{orcidlink}
\usepackage{wrapfig,framed,caption}

\usepackage[style=numeric-comp,sorting=none,maxbibnames=9,backend=bibtex]{biblatex}
\newcommand{\ellp}{\ell_\mathrm{P}}

\newcommand{\hr}{z_H}
\newcommand{\dho}{d_H}
\newcommand{\zh}{z_H}

\newcommand{\co}{z}
\newcommand{\ma}{\chi}

\newcommand{\kc}[1]{K_{#1}}
\newcommand{\sig}[1]{\sigma_{#1}}

\newcommand{\gen}{\mathcal{X}}
\newcommand{\wgen}{\widetilde{\mathcal{X}}}
\newcommand{\cgen}[1]{x_{#1}}
\newcommand{\wcgen}[1]{\widetilde{x}_{#1}}

\newcommand{\wtq}[1]{\widetilde{q}_{#1}}
\newcommand{\tp}[1]{\tau_{#1}}

\newcommand{\si}[1]{s_{#1}}
\newcommand{\sit}[1]{\widetilde{s}_{#1}}

\newcommand{\scheme}{EMD}

\newenvironment{linesmall}[1]
  {\arraycolsep=3pt\scriptsize
   \array{#1}}
  {\endarray}

\title{\LARGE \bf Effective Metric Descriptions\\ of \\  Quantum Black Holes}

\author[1,3,5]{Manuel Del Piano\thanks{{\Large \orcidlink{0000-0003-4515-8787}} \href{mailto:manuel.delpiano-ssm@unina.it}{manuel.delpiano-ssm@unina.it}}}
\author[2,4,5]{Stefan Hohenegger\thanks{{\Large \orcidlink{0000-0001-6564-0795}} \href{mailto:s.hohenegger@ipnl.in2p3.fr}{s.hohenegger@ipnl.in2p3.fr}}} 
\author[1,2,3,5]{Francesco Sannino\thanks{{\Large \orcidlink{0000-0003-2361-5326}} \href{mailto:sannino@qtc.sdu.dk}{sannino@qtc.sdu.dk}}}

\affil[1]{\small Scuola Superiore Meridionale, Largo S. Marcellino, 10, 80138 Napoli, Italy}
\affil[2]{\small Dept. of Physics E. Pancini, Università di Napoli Federico II, via Cintia, 80126 Napoli, Italy}
\affil[3]{\small INFN sezione di Napoli, via Cintia, 80126 Napoli, Italy}
\affil[4]{\small  Univ Lyon, Univ Claude Bernard Lyon 1, CNRS/IN2P3, IP2I Lyon, UMR 5822, F-69622, Villeurbanne, France }
\affil[5]{\small Quantum  Theory Center ($\hbar$QTC) \& D-IAS, Southern Denmark Univ., Campusvej 55, 5230 Odense M, Denmark}
\date{}

\begin{document}

\maketitle
\begin{abstract}
In a recent work \cite{DelPiano:2023fiw}, we have described spherically symmetric and static quantum black holes as deformations of the classical Schwarzschild {{metric}} that depend on the physical distance to the horizon. We have developed a framework that allows to compute the latter in a self-consistent fashion from the deformed {{geometry}}, in the vicinity of the horizon. However, in this formalism, the  distance can be replaced by other physical quantities, \emph{e.g.} curvature invariants such as the Ricci- or Kretschmann scalar. Here, we therefore define a more general framework, which we call an \emph{effective metric description} (\scheme), that captures the deformed geometry based on a generic physical quantity. We develop in detail the Ricci- and Kretschmann scalar \scheme, in particular demonstrating how to compute the geometry in a self-consistent   {{manner}}. Moreover, we provide explicit relations that allow to express one \scheme\,in terms of the others, thus demonstrating their equivalence.
\end{abstract}

\newpage
\tableofcontents
\section{Introduction}
In the past decades, considerable effort has been devoted to studying deformations and modifications of Einstein gravity. In particular, in the context of incorporating quantum effects, many different approaches are currently still being explored. Nevertheless, a comprehensive theory of quantum gravity is still missing. A recent approach  \cite{Binetti:2022xdi,DelPiano:2023fiw,DAlise:2023hls} (partially inspired by \cite{Bonanno:2000ep}) has therefore focused on describing (quantum) deformations of \emph{solutions} of General Relativity, namely black holes, in a universal  fashion: indeed, while for example a number of geometries \cite{bardeen1968proceedings,Hayward_2006,Dymnikova:1992ux,Donoghue:2001qc,Bjerrum-Bohr:2002fji,Kirilin:2006en,Calmet:2017qqa,Calmet:2019eof,Calmet:2021lny,Battista:2023iyu} have been proposed to capture deformations of the classical Schwarzschild space-time \cite{Schwarzschild}, our recent approach aims to capture model independent aspects of spherically symmetric and static quantum geometries. To this end, the introduction of a finite Planck scale is interpreted in the sense of the renormalisation group approach \cite{Barenblatt,barenblatt_1996,IntermediateAsymptotics,goldenfeld2018lectures,ChenGoldenfeldOono1995,oono1985advances} and non-trivial constraints stem from demanding that physical quantities are independent of spurious renormalisation scales. 

Concretely, the quantum deformations of the Schwarzschild metric are encoded in two functions that only depend on a \emph{physical quantity}, in order to guarantee that the modified geometry retains invariance under the same coordinate reparametrisations as its classical counterpart. Following the original idea in \cite{Bonanno:2000ep}, to this end, the physical \emph{distance} from the origin of the black hole was chosen. However, since this distance is calculated from the (deformed) metric, the geometry is not given explicitly, but only through the self-consistent solution of a complicated (and highly non-linear) differential equation. In \cite{DelPiano:2023fiw}, assuming sufficient regularity of the metric deformations, a framework was developed that allows to solve this self-consistency equation. Indeed, by expanding the problem in the vicinity (but outside of) the event horizon, exact and self-consistent solutions for the distance were found, which allow to calculate the metric to arbitrary order. Moreover, by computing physical quantities associated with the deformed space-time, notably the Ricci scalar (but also the Hawking temperature and the entropy), non-trivial conditions were found to ensure their finiteness at the horizon. These conditions were verified for certain models of quantum deformed black holes in the literature \cite{Hayward_2006,Dymnikova:1992ux}, while certain models were found to be incompatible \cite{Bonanno:2000ep}. 

Writing the metric deformations as a function of the physical distance to the black hole is a specific choice in \cite{DelPiano:2023fiw} and other physical quantities are indeed possible: for example, black hole deformations can be specified in terms of curvature invariants, notably the Ricci- and the Kretschmann scalar (see \emph{e.g.} the review \cite{Held:2021vwd} and references therein). This paper is devoted to explore spherically symmetric and static black hole solutions, which are self-consistently deformed (relative to the Schwarzschild geometry) with a single\footnote{In order to keep the discussion simple, we refrain from studying the most general deformation possible. The results found in this paper can, however, be generalised in a straight-forward fashion.} function depending on these invariants. We generalise the framework developed in \cite{DelPiano:2023fiw} for these cases and, assuming regularity and convergence, provide a solution of the respective self-consistency equations in the form of series expansions around the horizon. In doing so, we identify minimal sets of parameters that need to be specified to provide a complete description of the black hole geometry (at least in the vicinity of the horizon): these include expansion coefficients of the deformation functions as well as suitable boundary conditions for the self-consistency equations. We formalise this approach by grouping all this information into an \emph{effective metric description} (\scheme), which consists of 
\begin{itemize}
\item[\emph{(i)}] a physical quantity $\mathcal{X}$, like the physical distance or the Ricci scalar, which also lends its name to the \scheme; this also includes (normalisation) conditions and/or phase factors that characterise the classical limit of the geometry
\item[\emph{(ii)}] a set of {{free parameters}} required to specify (in a self-consistent fashion) the metric deformation
\item[\emph{(iii)}] a self-consistency equation which describes how $\mathcal{X}$ is computed from the metric
\end{itemize}
Following the distance \scheme\,used in \cite{DelPiano:2023fiw} as a blueprint, in this paper we also define the Ricci scalar- and Kretschmann scalar \scheme. Moreover, by expressing the metric deformations of the distance, Ricci scalar- and Kretschmann scalar-\scheme\,as series expansions in the Schwarzschild coordinate, we provide concrete relations that allow to express each \scheme\,in terms of the others. This demonstrates their equivalence and allows to explicitly express black hole solutions found in one \scheme\,in terms of another. The explicit relations among different \scheme s also highlight the fact that computations may be simpler in certain \scheme s. Indeed, while conceptually equivalent, \scheme s based on higher order curvature invariants typically require to solve differential equations that are non-linear in the deformation functions. This makes it difficult in certain cases to find solutions with certain properties, for example, simple limits for the classical black hole. This can be obtained more easily in \scheme s based on 'simpler' physical quantities.

Before continuing, we would like to clarify the relation of the current work (as an extension of \cite{Binetti:2022xdi,DelPiano:2023fiw,DAlise:2023hls}) compared to other approaches in the literature. Indeed, deformations of static and spherically symmetric black holes based on various physical quantities such as the distance (see \emph{e.g.}~\cite{Bonanno:2000ep,Platania:2023srt,Eichhorn:2022bgu}) or curvature scalars such as the Ricci- or Kretschmann-scalar (see \emph{e.g.}~\cite{Eichhorn:2021iwq,Eichhorn:2022bgu,Eichhorn:2022oma}) have been previously considered. However, rather than approximating these physical quantities by their classical counterparts \cite{Bonanno:2000ep,Eichhorn:2021iwq}\footnote{Indeed, we have shown in \cite{DelPiano:2023fiw} that such approximations can lead to non-physical behaviour, notably close to the horizon of the black hole.}, we develop a framework that allows to compute them exactly, in a self-consistent fashion, at least close to the horizon of the black hole. It is indeed the main result of our work that such frameworks (\emph{i.e.} \scheme s) can not only be developed for different physical quantities, but that they can also be uniquely transformed into one another. Furthermore, while our initial approach \cite{Binetti:2022xdi} is motivated by renormalisation group methods (following the general philosophy of \cite{Barenblatt,barenblatt_1996,IntermediateAsymptotics,goldenfeld2018lectures,ChenGoldenfeldOono1995,oono1985advances}) to find solutions of quantum black holes, the current work does not make use of this technology and therefore is different from other self-consistent approaches to find deformed black hole geometries, as for example \cite{Platania:2019kyx}: indeed, in the current work we simply provide a {\it standardised} framework to formulate (quantum) deformations of black holes in a physically meaningful manner and explain how different such frameworks can be interconnected.

Concretely, this paper is organised as follows: Section~\ref{Sect:ReviewDistance} serves the dual purpose of reviewing key results of previous work while at the same time illustrating our concept of an \scheme. Here the distance is our choice for the physical quantity. Sections~\ref{Sect:RicciScalar\scheme} and \ref{Sect:KretschmannScalar} introduce the Ricci- and Kretschmann scalar \scheme, providing in each case a solution of its self-consistency equation along with an explicit relation to the distance \scheme. Finally Section~\ref{Sect:Conclusions} contains a summary of our results along with our Conclusions. This work is complemented by 3 Appendices containing details on manipulating infinite series along with computational details for the Ricci- and Kretschmann scalar \scheme\,that were deemed too long to be included in the main body of the paper.

\section{Definitions and Review}
\subsection{Review of the Distance \scheme}\label{Sect:ReviewDistance}
This Section serves the dual purpose of reviewing part of our previous work~\cite{DelPiano:2023fiw}, while at the same time illustrating what we mean by an \scheme\, to describe (quantum) deformed Schwarzschild black holes. Our staring point is the metric of a spherically symmetric and static black hole of mass $\ma$ written in terms of the dimensionless coordinates\footnote{We prefer to work with dimensionless quantities such that $\dd s$, $t$ and $\co$ correspond to the infinitesimal line element, the time coordinate and the radial Schwarzschild coordinate, respectively, measured in units of the Planck length $\ellp$. Similarly, the mass $\ma$ is measured in units of the Planck mass $M_\mathrm{P}=1/\ellp$. We refer to previous work \cite{Binetti:2022xdi,,DelPiano:2023fiw} for more discussion of this point.} $(t,\co,\theta,\varphi)$ 
\begin{align}
&\dd s^2=-f(\co) \dd t^2+\frac{\dd \co^2}{f(\co)}+\co^2 \dd \theta^2+\co^2 \sin^2\theta \, \dd\varphi^2 \qq{with} f(\co)=1-\frac{2\ma}{\co}\,v^{(\mathcal{X})}\,.\label{SpStMetric}
\end{align}
We assume that $f(\co)$ has a simple zero at some finite $\zh>0$, which is the position of the (outer) event horizon of the black hole. In the following we shall discuss (\ref{SpStMetric}) only for $\co\geq\zh$ and we assume $f$ to be positive definite in this region. We also remark that here we do not consider the most general static space-time metric compatible with spherical symmetry, but to streamline the discussion, we consider a family of metrics that are parametrised by a single function $v^{(\mathcal{X})}$ (see \cite{DelPiano:2023fiw} for a discussion of the most general metric) {{of a single variable.}}  An \scheme\,is largely determined by identifying the latter with a physical quantity $\mathcal{X}$, which is invariant under reparametrisations and which is a monotonic function of the Schwarzschild coordinate $\co$.\footnote{We shall often use the (slightly) abusive notation $v^{(\mathcal{X})}(\co)$ to signify $v^{(\mathcal{X})}(\mathcal{X}(\co))$. Furthermore, since the goal of this paper is to compare \scheme s defined by different physical quantities, we shall encounter various different deformation functions and the superscript $(\mathcal{X})$ simply serves as a naming convention.} To explain this in more detail, we follow the example of \cite{Binetti:2022xdi,DelPiano:2023fiw} and choose this physical quantity to be the distance $\rho$ measured from the horizon of the black hole, \emph{i.e.} $v^{(\rho)}(\rho)$. In writing $f(\co)$ in (\ref{SpStMetric}), we implicitly understand $\rho$ to be a (monotonic) function of the Schwarzschild coordinate $\co$, which is fixed through the first order differential equation
\begin{align}
&\frac{\dd\rho}{\dd\co}=\frac{1}{\sqrt{f}}\,,&&\text{and}&&\rho(\co=\zh)=0\,.\label{SelfEquationDistance}
\end{align}
This differential equation is non-linear, since the right hand side through $v^{(\rho)}$ is also a function of the distance $\rho$. In \cite{DelPiano:2023fiw} a framework was proposed to solve this equation in a self-consistent fashion as a series expansion close to the horizon. To this end, the following series expansion was defined
\begin{align}
v^{(\rho)}(\rho)=\frac{1}{2\chi}\,\sum_{n=0}^\infty \xi_n\,\rho^n\,&&\text{with}&&\xi_n\in\mathbb{R}\hspace{0.2cm}\forall n\geq0\,,&&\text{and} &&\xi_0=\zh\,,\label{SeriesDistanceRho}
\end{align}
such that the deformation of the classical Schwarzschild geometry is fully encoded in the coefficients $\{\xi_n\}_{n\in\mathbb{N}^*}$. The series in (\ref{SeriesDistanceRho}) is assumed to have a non-vanishing radius of convergence { {(see \cite{ruse1931,synge1931,dewitt1960radiation} where a Taylor theorem for more general tensor calculus, based on (various) geodesic distances, was developed)}}. Furthermore, self-consistency (concretely the fact that all derivatives of $f$ with respect to the Schwarzschild variable $\co$ are well-defined at the horizon \cite{DelPiano:2023fiw}) requires $\xi_{2n-1}=0$ $\forall n\in\mathbb{N}$ and $\xi_2\leq \frac{1}{16\zh}$, which we shall assume to hold in the following. Using the expansion (\ref{SeriesDistanceRho}) allows to express the distance $\rho$ as a function of the Schwarzschild coordinate $\co$ and vice-versa
\begin{align}
&\co(\rho)=\zh+\sum_{n=1}^\infty a_n\,\rho^n\,,&&\text{and} &&\rho(\co)=\sum_{n=1}^\infty b_n\,(\co-\zh)^{n/2}\,.\label{SeriesDistance}
\end{align}
Here the coefficients $a_n$ are expressed in terms of the $\xi_n$ as follows 
\begin{align}
&a_1=0\,,\hspace{1cm}a_2=\frac{1+v\varpi}{8\,\zh}\,,\hspace{1cm}\text{with} \hspace{1cm}\varpi=\sqrt{1-16\,\zh\, \xi_2}\,,\nonumber\\
&a_p=\frac{\xi_p+\zh\sum_{n=3}^{p-1}(p-n+2)\,n\,a_n\,a_{p-n+2}+\sum_{n=2}^{p-2}\sum_{m=2}^n(n-m+2)\,m\,a_{p-n}\,a_m\,a_{n-m+2}}{1-4\zh\, p\, a_2}\,,\label{Coeffsa}
%
\end{align}
and $b_n$ follow from inversion of the series (\ref{SeriesDistance})\footnote{Notice that for $\xi_2\leq\frac{1}{16\zh}$, the coefficient $a_2\neq 0$.}
\begin{align}
&b_1=\frac{1}{a_2^{1/2}}\,,&&b_2=-\frac{a_3}{2a_2^2}\,,&&b_n=\frac{(-1)^{n-1}}{2^{n-1}\,n!\, a_2^{n/2}}\,\text{det}(\mathcal{M}_n)\hspace{0.5cm}\forall n\geq 3\,.\label{Coeffsb}
\end{align}
$v=\pm 1$ in (\ref{Coeffsa}) is an ambiguous sign parameter which arises from solving (\ref{SelfEquationDistance}). In \cite{DelPiano:2023fiw} $v=+1$ was chosen, which is a necessary condition for the black hole to reproduce the Schwarzschild space-time in the limit $\xi_n\to 0$ $\forall n\in\mathbb{N}$. For simplicity, in the following we shall consider the same choice, which, however, we consider as an additional condition to define the \scheme. The conditions $\xi_{2n-1}=0$ $\forall n\in\mathbb{N}$ imply
\begin{align}
&a_{2n-1}=0=b_{2n}\,,&&\forall n\in\mathbb{N}\,.\label{ABrestriction}
\end{align}
Furthermore, the $(n-1)\times (n-1)$ matrix $\mathcal{M}_n$ in (\ref{Coeffsb}) is explicitly given in \cite{DelPiano:2023fiw}, but its details shall not be important in the following: we only remark that one can write
\begin{align}
&b_{2n-1}=\frac{2^{3n}\sqrt{2\frac{1+\varpi}{\zh}}\zh^{n+1}\,\xi_{2n}}{(1+\varpi)^{n+1}((n-1)+n\varpi)}+\widetilde{b}_{2n-1}(\xi_2,\ldots,\xi_{2n-2})\,,&&\forall n\geq 2\,,\label{ConditionExpansionBs}
\end{align}
where notably $\widetilde{b}_{2n-1}$ is independent of $\xi_{2k}$ for $k>n$.

To summarise and formalise the discussion, we characterise an \scheme\,for describing a (quantum) deformed spherically symmetric and static black hole by the following 
\begin{wrapfigure}{r}{0.25\textwidth}
\begin{center}
\scalebox{1}{\parbox{4.2cm}{\begin{tikzpicture}        
\draw[rounded corners,fill=cyan!45!white] (-2.1, -0.1) rectangle (2.1, -3.1);
\node at (0,-0.5) {\bf Distance $\rho$};
\node at (-0.1,-1) {\tiny free parameters (\ref{SeriesDistanceRho}): $\{\xi_{2n}\}_{n\in\mathbb{N}^*}$};
\node at (-0.3,-1.4) {\tiny self-consistency equation (\ref{SelfEquationDistance}):};
\node at (0,-1.9) {$\frac{\dd\rho}{\dd\co}=\frac{1}{\sqrt{f}}$};
\node at (-0.1,-2.4) {\tiny  normalisation condition (\ref{SelfEquationDistance}), (\ref{Coeffsa}):};
\node at (0,-2.8) {\tiny $\rho(\zh)=0$, $v=+1$, $\xi_2\leq\frac{1}{16\zh}$};
\end{tikzpicture}
}}
\caption{Summary of the distance \scheme.}
\label{Fig:SummaryDistance}
\end{center}
${}$\\[-2cm]
\end{wrapfigure}
\noindent
set of quantities:
\begin{enumerate}
\item[\emph{(i)}] a defining {\bf physical quantity} $\mathcal{X}$; We shall mostly use $\mathcal{X}$ to loosely label the \scheme, \emph{e.g.} we shall call the \scheme\,outline above simply the \emph{distance}-\scheme. Furthermore, additional {\bf (normalisation) condition(s)} may be required that supplement (and sharpen) the definition of the physical quantity: these can act as boundary conditions for the self-consistency condition (point \emph{(iii)} below), or are related to the classical limit (in which the metric deformation is switched off). In the example of the distance \scheme, this condition is the convention to measure the distance from the horizon of the black hole \emph{i.e.} $\rho(\zh)=0$ as well as the choice of the sign $v=+1$ in (\ref{Coeffsa}) (which allows for the recovery of the Schwarzschild geometry in the limit $\xi_{2n}\to 0$ $\forall n\geq 1$).   
\end{enumerate}

\begin{enumerate}
\item[\emph{(ii)}] a {\bf set of {{free parameters}}} which are used to deform the metric function $f$ in the metric (\ref{SpStMetric}); Usually, we refer more concretely to expansion coefficients of $v^{(\mathcal{X})}$ close to the horizon (as \emph{e.g.} in (\ref{SeriesDistanceRho}))
\item[\emph{(iii)}] a {\bf self-consistency equation} which encodes how $\mathcal{X}$ is computed from the metric; This equation usually takes the form of a (non-linear) differential equation (see (\ref{SelfEquationDistance}) in the case of the distance \scheme).
\end{enumerate}
These elements are briefly summarised in Figure~\ref{Fig:SummaryDistance} for the distance \scheme\,(\emph{i.e.} for $\mathcal{X}=\rho$). Once the \scheme\, has been defined, physical quantities related to the black hole geometry, can be computed in terms of the {{free parameters}}. For example, in \cite{DelPiano:2023fiw} the Hawking temperature was computed as the surface gravity \cite{Bogoliubov1958,Wald:1984rg,Sarkar:2007uz} of the black hole \cite{Hawking1975,birrell_davies_1982}\footnote{This expression takes into account that here we only consider a single metric deformation function.}
\begin{align}
T_H=\frac{1}{4\pi}\,\frac{\dd f}{\dd \co}\bigg|_{\co=\zh}=\frac{1+\varpi}{8\pi\zh}\,,\label{HawkingTempFirst}
\end{align}
which is indeed only a function of $\varpi$ and $\zh$ (or equivalently $\xi_{0,2}$).

Before closing this Subsection, we make a few remarks concerning the work \cite{DelPiano:2023fiw}: in addition to (\ref{SeriesDistanceRho}) the deformation function in \cite{DelPiano:2023fiw} has also been called $v^{(\rho)}=e^{\Phi\left(\frac{1}{\dho+\rho}\right)}$, where $\Phi:\,\mathbb{R}_+\to\mathbb{R}$ is a function (which is (several times) differentiable at the horizon) and $\dho$ is the distance of the horizon to the center of the black hole. Written in this way, the deformation is rather seen as a function of the distance $d(z)=\dho+\rho(z)$ to the origin. As we shall explain in the next Subsection, changing the \scheme\,by shifting the physical quantity $\mathcal{X}$ by a constant (in this case $\dho$) is a simple \scheme-redefinition, which can be performed in a straight-forward manner (assuming that this shift does not change drastically the convergence properties of (\ref{SeriesDistanceRho})). Notice, however, that using $e^{\Phi\left(\frac{1}{\dho+\rho}\right)}$ requires the knowledge of $\dho$ (which in \cite{DelPiano:2023fiw} was taken as additional input), which in (\ref{SeriesDistanceRho}) is implicitly encoded in the coefficients $\{\xi_n\}_{n\in\mathbb{N}^*}$.

Another important point concerns the fact that (\ref{SeriesDistanceRho}) as well as (\ref{SeriesDistance}) are in fact (infinite) series. As we have already mentioned, we assume that these series have a finite radius of convergence (\emph{i.e.} (\ref{SeriesDistanceRho}) as well as (\ref{SeriesDistance}) are well defined at least sufficiently close to the horizon). This also implicitly requires that the deformation function $v^{(\rho)}$ (and thus by extension also the metric function $f(z)$) are infinitely many times   {{differentiable}} at the horizon. In \cite{DelPiano:2023fiw} we have {{also}} derived {{weaker}} consistency conditions for $v^{(\rho)}$ (more concretely for (derivatives of) the function $\Phi$ mentioned above) requiring \emph{only} the existence of the first and second derivative of $f$ with respect to the Schwarzschild coordinate. In the current work, however, we indeed require existence of the full series expansion (\ref{SeriesDistanceRho}) as well as (\ref{SeriesDistance}) (\emph{i.e.} infinitely many terms) in order to (fully) define the \scheme\,and to be able to connect it to other \scheme s we shall introduce below. In other words, we assume an \scheme\,to  be fixed by knowing the entire function $v^{(\mathcal{X})}$, which here we consider encoded in infinitely many expansion coefficients.

\subsection{\scheme\,redefinitions}\label{Sect:ShiftOfFunction}
Before discussing \scheme s  based on different physical quantities (that is curvature scalars rather than the physical distance), we shall first consider \scheme\,redefinitions, in which the physical quantity $\gen$ is replaced by simple functions of it: indeed, consider the starting point 
\begin{align}
v^{(\gen)}(\gen)=\sum_{n=0}^\infty \cgen{n}\,\gen^n\,&&\text{with}&&\cgen{n}\in\mathbb{R}\hspace{0.2cm}\forall n\geq0\,,\label{SeriesGenPhys}
\end{align}
which we assume to have finite radius of convergence $\gen_c$, and replace $\gen$ by a continuous, monotonic function of $\gen$, {{which we denote}} $\sigma:\,[0,\gen_c]\longrightarrow \mathbb{R}$. Then we define $\wgen=\sigma(\gen)$, along with the deformation function
\begin{align}
v^{(\wgen)}(\wgen)=\sum_{n=0}^\infty \wcgen{n}\,\wgen^n\,&&\text{with}&&\wcgen{n}\in\mathbb{R}\hspace{0.2cm}\forall n\geq0\,.\label{SeriesGenPhysMod}
\end{align}
A very simple example of this idea was already used in \cite{DAlise:2023hls}, where a physical distance shifted by a constant was considered, \emph{i.e.} $\widetilde{\gen}=\sigma(\gen)=\gen+c$ for $c\in\mathbb{R}$. In this case, the two sets of coefficients $\{\cgen{n}\}_{n\in\mathbb{N}^*}$ and $\{\wcgen{n}\}_{n\in\mathbb{N}^*}$ are related
\begin{align}
&\cgen{n}=\sum_{\ell=n}^\infty\frac{\wcgen{\ell}\,c^{\ell-n}}{(\ell-n)!}\,,&&\wcgen{n}=\sum_{\ell=n}^\infty\frac{\cgen{\ell}\,(-c)^{\ell-n}}{(\ell-n)!}\,,&&\forall n\in\mathbb{N}\,.\label{ConstShiftRel}
\end{align}
Notice that in this case, any single $\cgen{n}$ depends on infinitely many of the $\{\wcgen{n'}\}_{n'\geq n}$ and vice-versa. 

In order to describe the relation between (\ref{SeriesGenPhys}) and (\ref{SeriesGenPhysMod}) for a more general function $\sigma$, we assume that the latter can be expanded 
\begin{align}
&\sigma(\gen)=\sum_{k=1}^\infty \sig{k}\,\gen^k\,,&&\text{with} &&\sig{k}\in\mathbb{R}\,,\label{SeriesExpandL}
\end{align}
which has a radius of convergence $\gen'_c\geq \gen_c$.\footnote{Since (\ref{ConstShiftRel}) already describes the relation for a shift by a constant, we have assumed $\sig{0}=0$. Furthermore, in the following we shall implicitly assume $\sig{1}\neq 0$.}
By demanding $v^{(\gen)}(\gen)=v^{(\wgen)}(\wgen)$
\begin{align}
v^{(\wgen)}(\wgen)=\sum_{n=0}^\infty \wcgen{n}\,\left(\sum_{k=1}^\infty \sig{k}\,\gen^k\right)^n
=v^{(\gen)}(\gen)\,,
\end{align}
and using (\ref{DefRedFormPower}) in Appendix~\ref{App:PowerPower}, we obtain the following relation
\begin{align}
&\cgen{0}=\wcgen{0}\,,\hspace{1cm}\cgen{1}=\sig{1}\,\wcgen{1}\,,\nonumber\\
&\cgen{m}=\sig{m}\,\wcgen{1}+\sum_{p=2}^m \wcgen{p}   \sum_{{0\leq \ell_{p-1}\leq \ell_{p-2}\leq \ldots\leq \ell_1\leq n}\atop{\ell_1+\ldots+\ell_{p-1}=p-n}}\,\frac{n!\,\sig{1}^{n-\ell_1}\ldots \sig{p-1}^{\ell_{p-2}-\ell_{p-1}}\sig{p}^{\ell_{p-1}}}{(n-\ell_1)!\ldots (\ell_{p-2}-\ell_{p-1})!\ell_{p-1}!}\,.\label{FormOmegRel}
\end{align}
This identification is a linear relation, which expresses $\cgen{m}$ as a linear combination of \emph{finitely} many $\wcgen{n}$ with $n\leq m$. Formally, this relation can be written as a linear transformation with an infinite dimensional lower triangular matrix
\begin{align}
\left(\begin{array}{c}\cgen{0} \\ \cgen{1} \\ \cgen{2} \\ \cgen{3} \\ \cgen{4} \\ \cgen{5}\\ \vdots\end{array}\right)=\left(\begin{array}{ccccccc}1 & 0 & 0 & 0 & 0& 0 & \cdots \\ 0 & \sig{1} & 0 & 0 & 0 & 0 & \cdots \\ 0 & \sig{2} & \sig{1}^2 & 0 & 0 & 0 & \cdots \\ 0 & \sig{3} & 2\sig{1}\sig{2} & \sig{1}^3 & 0 & 0 & \cdots \\ 0 & \sig{4} & \sig{2}^2+2\sig{1}\sig{3} & 3\sig{1}^2 \sig{2} & \sig{1}^4 &  0 & \cdots \\ 0 & \sig{5} & 2(\sig{1}\sig{4}+\sig{2}\sig{3}) & 3(\sig{1}\sig{2}^2+\sig{1}^2\sig{3}) & 4\sig{1}^3 \sig{2} & \sig{1}^5 & \cdots\\ \vdots & \vdots & \vdots & \vdots &\vdots & \vdots & \ddots\end{array}\right)\cdot\left(\begin{array}{c}\wcgen{0} \\ \wcgen{1} \\ \wcgen{2} \\ \wcgen{3} \\ \wcgen{4} \\ \wcgen{5}\\ \vdots\end{array}\right)\,.\label{GenLinTransform}
\end{align}
Combining (\ref{ConstShiftRel}) with (\ref{GenLinTransform}) allows to describe more general \scheme\,redefinitions. 
 
\section{Ricci Scalar \scheme}\label{Sect:RicciScalar\scheme}
In this Section we shall discuss a new \scheme\,that is not based on the physical distance to the black hole (horizon), but instead to the Ricci scalar. Indeed, outside of the horizon of the black hole, the latter is expected to be a monotonic function of the Schwarzschild coordinate $\co$. Here we shall first discuss how to define such deformations in a self-consistent fashion and then explain how such descriptions can be linked to the distance \scheme\,that we have reviewed in the previous Section.

\subsection{Definition of the \scheme}
To define the Ricci scalar \scheme, we shall use for the deformation of the metric (\ref{SpStMetric}) a function $v^{(R)}$ of the Ricci scalar:
\begin{align}
&f(\co)=1-\frac{2 \ma}{\co}\,v^{(R)}(R(\co))\,.
\end{align}
As before, we consider the outer horizon of the black hole to be located at $\co=\hr$, such that $f(\hr)=0$. The Ricci scalar $R$ is fixed in terms of the metric function in the following manner\footnote{In this work, primes denote derivatives with respect to the Schwarzschild coordinate $\co$. Furthermore we have used the shorthand notation $\frac{\dd v^{(R)}}{\dd \co}=\frac{\dd v^{(R)}}{\dd R}\,\frac{\dd R}{\dd \co}$, and similarly for the second derivative.}
\begin{align}
R(\co)=\frac{1}{\co^2}\,\left(2-2f -4\co f'-\co^2 f''\right)=\frac{2\ma}{\co^2} \left(2\dv{v^{(R)}}{\co}+\co \dv[2]{v^{(R)}}{\co}\right)\,,\label{FormRicciEq}
\end{align} 
which is a second order, non-linear differential equation for $R$ as a function of the Schwarzschild coordinate $\co$. Solving (\ref{FormRicciEq}) in general in the regime $\co\geq \zh$ is a difficult task and we limit ourselves to studying a series expansion in the vicinity of the horizon. Concretely, we define the following series expansions
\begin{align}
&v^{(R)}(R)=\sum_{n=0}^\infty \frac{q_n}{n!}\,R^n\,,&&\text{and} &&  R(\co)=\sum_{n=0}^\infty \frac{R_n}{n!}\,(\co-\hr)^n\, ,\label{SeriesExpansionszh}
\end{align}
which we assume to have a non-zero radius of convergence for $z\geq \zh$. Notice that the sets of expansion coefficients $\{q_n\}_{n\in\mathbb{N}^*}$ and $\{R_{n}\}_{n\in\mathbb{N}^*}$ play very different roles: the $\{q_n\}$ implicitly define the function $v^{(R)}$ and are therefore external input parameters that \emph{define} the metric deformation. They are therefore part of the {{free parameters}} that define the Ricci scalar \scheme. The coefficients $\{R_n\}$ are a priori not free, but instead fixed by (\ref{FormRicciEq}) in terms of the $\{q_n\}$. However, since (\ref{FormRicciEq}) is a second order differential equation, its general solution depends on two further (unfixed) parameters. Thus, in order to fully describe the Ricci scalar \scheme, we first need to understand how to parametrise this freedom.

First of all, the condition of the horizon being located at $\zh$ (\emph{i.e.} $f(\hr)=0$) imposes
\begin{align}
\sum_{n=0}^\infty\frac{q_n}{n!}\,(R_0)^n=\frac{\hr}{2\chi}\ .
\end{align}
Furthermore, the expansion of $R$ in (\ref{SeriesExpansionszh}) allows to expand $v^{(R)}$ in terms of the Schwarzschild coordinate 
\begin{align}\label{v exp r}
v^{(R)}(R)=\sum_{n=0}^\infty \frac{q_n}{n!}\,\left(\sum_{k=0}^\infty \frac{R_k}{k!}\,(\co-\hr)^k\right)^n=:\sum_{p=0}^\infty \tp{p}\,(\co-\hr)^p\ .
\end{align}
This series can be developed in the following form (see Appendix~\ref{App:PowerPower} for a discussion on the technical details of expanding powers of series)
\begin{align}
v^{(R)}&=\sum_{n=0}^\infty \frac{q_n}{n!}\sum_{\ell_1=0}^n\sum_{\ell_2=0}^{\ell_1}\ldots\sum_{\ell_p=0}^{\ell_{p-1}}\begin{pmatrix} n \\ \ell_{1}\end{pmatrix} \begin{pmatrix} \ell_{1} \\ \ell_{2} \end{pmatrix} \ldots \begin{pmatrix} \ell_{p-1} \\ \ell_{p}\end{pmatrix}\,(R_0)^{n-\ell_1}\,\left(\tfrac{R_1}{1!}\right)^{\ell_1-\ell_2}\ldots \left(\tfrac{R_{p-1}}{(p-1)!}\right)^{\ell_{p-1}-\ell_p} \times \nonumber\\
&\hspace{1cm} \times \left(\tfrac{R_p}{p!}\right)^{\ell_p}\,(\co-\hr)^{\ell_1+\ell_2+\ldots \ell_{p-1}+\ell_p}+\mathcal{O}((\co-\hr)^{p+1})\ ,
\end{align}
such that we can read off the coefficients
\begin{align}\label{cp coeff}
\tp{0}&=\sum_{n=0}^\infty \frac{q_n}{n!}\,(R_0)^n\,,\nonumber\\
\tp{p}&=\sum_{n=0}^\infty q_n\,\sum_{{0\leq \ell_p\leq \ell_{p-1}\leq\ldots\leq \ell_1\leq n}\atop{\ell_1+\ell_2+\ldots+\ell_p=p}}\frac{(R_0)^{n-\ell_1}\,\left(\tfrac{R_1}{1!}\right)^{\ell_1-\ell_2}\ldots \left(\tfrac{R_{p-1}}{(p-1)!}\right)^{\ell_{p-1}-\ell_p}\left(\tfrac{R_p}{p!}\right)^{\ell_p}}{(n-\ell_1)!\,(\ell_1-\ell_2)!\ldots (\ell_{p-1}-\ell_p)!\ell_p!}\,,\hspace{0.25cm} \forall p\geq 1\,.
\end{align}
Furthermore, using the expansions \cite{DelPiano:2023fiw}
\begin{align}
&\frac{1}{\co}=\sum_{n=0}^\infty\frac{(-1)^n}{\hr^{n+1}}\,(\co-\hr)^n \,,&&\text{and} && \frac{1}{\co^2}=\sum_{n=0}^\infty\frac{(-1)^n(n+1)}{\hr^{n+2}}\,(\co-\hr)^n \ ,
\end{align}
we can further expand the two terms on the right hand-side of eq.~\eqref{FormRicciEq} by using the Cauchy product of two series
\begin{align}
\frac{1}{\co^2}\,\dv{v^{(R)}}{\co}&=\sum_{p=0}^\infty (\co-\hr)^{p}\sum_{n=0}^p\frac{(-1)^n(n+1)(p+1-n)}{\hr^{n+2}}\,\tp{p+1-n}\ ,\\
\frac{1}{\co}\,\dv[2]{v^{(R)}}{\co}&=\sum_{p=0}^\infty(\co-\hr)^p\sum_{n=0}^p\frac{(-1)^n(p+2-n)(p+1-n)}{\hr^{n+1}}\,\tp{p+2-n}\ .
\end{align}
Combining these results, (\ref{FormRicciEq}) can be expanded in powers of $(\co-\hr)$
\begin{align}
\sum_{p=0}^\infty\frac{R_p}{p!}\,(\co-\hr)^p&=2\chi\sum_{p=0}^\infty(\co-\hr)^p\sum_{n=0}^p\frac{(-1)^n}{\hr^{n+2}}\big[2\hr(n+1)(p+1-n)\,\tp{p+1-n}+\nonumber\\
&\hspace{5cm}+(p+2-n)(p+1-n)\,\tp{p+2-n}\big]\,,\label{LinRelRp}
\end{align}
such that, by comparing order by order, we find
\begin{align}
\frac{R_p}{p!}=2\chi\sum_{n=0}^p\frac{(-1)^n}{\hr^{n+2}}\left[2(n+1)(p+1-n)\,\tp{p+1-n}+\hr(p+2-n)(p+1-n)\,\tp{p+2-n}\right]\ .\label{RecursiveRp}
\end{align}
To understand this relation, we recall that the $\tp{p}$ in (\ref{cp coeff}) are functions of the coefficients $\{q_n\}$ and $\{R_n\}$. Therefore, a priori (\ref{RecursiveRp}) allows to express the latter in terms of the former. Concretely, however, (\ref{RecursiveRp}) can be seen as a recursive equation that allows to express iteratively the $\{R_n\}_{n\geq 2}$ in terms of $\{R_0,R_1,q_{n\geq 0}\}$. To see this, we re-write the second equation of \eqref{cp coeff} in the form
\begin{align}
\tp{p+2}=\frac{R_{p+2}}{(p+2)!}\sum_{n=1}^\infty\frac{q_n\,R_0^{n-1}}{(n-1)!}+\tau^{\text{sub-}R}_{p+2}(R_0,\ldots,R_{p+1},\{q_n\})\ ,\label{Leading1}
\end{align}
where $\tau^{\text{sub-}R}_{p+2}$ is notably independent of $R_{k\geq p+2}$
\begin{align}\label{tilde cp+2}
\tau^{\text{sub-}R}_{p+2}:=\sum_{n=0}^\infty q_n\,\sum_{{0\leq \ell_{p+1}\leq\ldots\leq \ell_1\leq n}\atop{\ell_1+\ell_2+\ldots\ell_{p+1}=p+2}}\frac{R_0^{n-\ell_1}\,\left(\tfrac{R_1}{1!}\right)^{\ell_1-\ell_2}\ldots \left(\tfrac{R_{p+1}}{(p+1)!}\right)^{\ell_{p+1}}}{(n-\ell_1)!\,(\ell_1-\ell_2)!\ldots (\ell_{p}-\ell_{p+1})!\ell_{p+1}!}\, .
\end{align} 
Therefore (\ref{RecursiveRp}) is a linear equation in $R_{p+2}$ that allows to express it in terms of  $R_{p'}$, with $p^{\prime}<p+2$ and $\{q_n\}$
\begin{align}\label{Rp+2 iterative}
&R_{p+2}=\frac{\hr p!}{2\chi\sum_{k=1}^\infty\frac{q_k R_0^{k-1}}{(k-1)!}}\bigg[\frac{R_p}{p!}-\frac{2\ma (p+2)(p+1)}{\hr}\,\tau^{\text{sub }R}_{p+2}+\nonumber\\
&-4\chi\sum_{n=0}^p\frac{(-1)^n(n+1)(p+1-n)}{\hr^{n+2}}\tp{p+1-n}-2\chi\sum_{n=1}^p\frac{(-1)^n(p+2-n)(p+1-n)}{\hr^{n+1}}\,\tp{p+2-n}\bigg]\ .
\end{align}
Thus, we can choose as the {{free parameters}} to describe this \scheme\,the parameters $R_0$, $R_1$

\begin{wrapfigure}{l}{0.27\textwidth}
${}$\\[-1cm]
\begin{center}
\scalebox{1}{\parbox{4.2cm}{\begin{tikzpicture}        
\draw[rounded corners,fill=green!45!white] (-2.1, -0.1) rectangle (2.1, -3.1);
\node at (0,-0.5) {\bf Ricci scalar $R$};
\node at (-0.05,-1) {\tiny param. (\ref{SeriesExpansionszh}): $R_0\,,R_1\,,\{q_n\}_{n\in\mathbb{N}^*}$};
\node at (-0.2,-1.4) {\tiny self-consistency equation (\ref{FormRicciEq}):};
\node at (0,-1.9) {\tiny $R=\frac{1}{\co^2}\,\left(2-2f -4\co f'-\co^2 f''\right)$};
\node at (-0.5,-2.4) {\tiny  normalisation condition:};
\node at (0,-2.8) {{\tiny $R(\zh)=R_0$}};
\end{tikzpicture}
}}
\caption{Summary of the Ricci scalar \scheme.}
\label{Fig:SummaryRicci}
\end{center}
${}$\\[-1cm]
\end{wrapfigure}
\noindent
and $\{q_n\}_{n\in\mathbb{N}^*}$. The discussion so far is summarised in Figure~\ref{Fig:SummaryRicci}, which formalises the definition of the Ricci scalar \scheme\,to describe (quantum) deformations of a spherically symmetric and static black hole. Notice that this \scheme\,a priori gives a different description of the black hole deformation as the distance \scheme\,discussed in the previous Section. Moreover, given the different sets of input parameters (\emph{i.e.} $\{\xi_n\}_{n\in\mathbb{N}^*}$ for the distance \scheme\,and $(R_0,R_1,\{q_n\})$ for the Ricci scalar \scheme) it is not clear whether these two descriptions are equivalent. In the following Subsection we shall therefore discuss how to connect the two \scheme s, that is, we shall try to formulate relations among the two sets of parameters that lead to identical deformations and therefore the same geometry.\footnote{Notice that we shall impose that the two deformations are indeed \emph{identical}. We shall not discuss the much more general possibility of the two space-times described by these deformations being equivalent and related to one another through coordinate transformations.}

\subsection{Relation to the Distance \scheme}
Our strategy in relating the distance \scheme\,to the Ricci scalar \scheme\,is to identify the deformation functions, \emph{i.e.} $v^{(R)}=v^{(\rho)}$. This can for example be achieved by expanding both functions in the Schwarzschild coordinate and identify the coefficients $\tp{p}$ (as defined in (\ref{v exp r})) order by order. However, a major technical complication on the Ricci scalar side is the fact that $\tp{p}$ for fixed $p$ generally depends on all $\{q_n\}_{n\in\mathbb{N}^*}$. We therefore first slightly re-define the Ricci scalar \scheme\,before relating it to the distance \scheme.

\subsubsection{Redefinition of the Ricci Scalar \scheme}
We may slightly re-define the Ricci scalar \scheme\,by choosing as physical quantity $\widetilde{R}(z):=R(z)-R_0$. Since $R_0$ is a constant (namely the value of the Ricci scalar at the horizon), $\widetilde{R}$ is a physical quantity and (outside of the horizon) a monotonic function of $z$. We thus consider the expansion of the modified deformation function
\begin{align}
&v^{(\widetilde{R})}=\sum_{n=0}^\infty\frac{\wtq{n}}{n!}\,\widetilde{R}^n\,,&&\text{and} &&\widetilde{R}(\co)=\sum_{n=1}^\infty\frac{R_n}{n!}\,(\co-\zh)^n\,.\label{SeriesExpansionRtzh}
\end{align}
Following the same logic as explained in Section~\ref{Sect:ShiftOfFunction} (see (\ref{ConstShiftRel})), the coefficients $q_p$ in (\ref{SeriesExpansionszh}) and the $\wtq{p}$ in (\ref{SeriesExpansionRtzh}) can be related through
\begin{align}
&q_p=\sum_{n=p}^\infty\frac{\wtq{n}\,(-R_0)^{n-p}}{(n-p)!}\,,&&\text{and}&&\wtq{p}=\sum_{n=p}^\infty\frac{q_n\,R_0^{n-p}}{(n-p)!}\,.\label{SchemeReductionR}
\end{align}
Furthermore, the identification (\ref{SchemeReductionR}) allows to write
\begin{align}
v^{(\widetilde{R})}(\widetilde{R})=\sum_{p=0}^\infty \tp{p}\,(\co-\zh)^p\,, \label{ExpansionDeformationFunction}
\end{align}
with the same coefficients $\tp{p}$ as in (\ref{v exp r}). Concretely, we find the following form for $\tp{p}$ as a function of $\{\wtq{n}\}_{n\in\mathbb{N}^*}$ and $\{R_n\}_{n\in\mathbb{N}}$
\begin{align}
\tp{0}&=\wtq{0}\,,\nonumber\\
\tp{p}&=\sum_{{0\leq \ell_{p-1}\leq\ldots\leq \ell_1\leq n\leq p}\atop{\ell_1+\ell_2+\ldots+\ell_{p-1}+n=p}} \frac{\wtq{n}\,\left(\tfrac{R_1}{1!}\right)^{n-\ell_1}\left(\tfrac{R_2}{2!}\right)^{\ell_1-\ell_2}\ldots \left(\tfrac{R_{p-1}}{(p-1)!}\right)^{\ell_{p-2}-\ell_{p-1}}\left(\tfrac{R_{p}}{p!}\right)^{\ell_{p-1}}}{(n-\ell_1)!\,(\ell_1-\ell_2)!\ldots (\ell_{p-2}-\ell_{p-1})!\ell_{p-1}!}\hspace{0.5cm} \forall p\geq 1\,.\label{tpCoefsForm}
\end{align}
Inserting this relation in the self-consistency relation (\ref{RecursiveRp}) leads to a recursive relation that fixes $R_n$ for $n\geq 2$ in terms of $R_0$, $R_1$ and $\{\wtq{n}\}_{n\in \mathbb{N}^*}$:
\begin{align}\label{Rp+2 iterative 2}
&R_{p+2}=\frac{\hr p!}{2\chi\wtq{1}}\bigg[\frac{R_p}{p!}-\frac{2\ma(p+2)(p+1)}{\hr}\,\sum_{{0\leq \ell_{p}\leq\ldots\leq \ell_1\leq n\leq p+2}\atop{\ell_1+\ell_2+\ldots\ell_{p}+n=p+2}}\frac{\wtq{n}\left(\tfrac{R_1}{1!}\right)^{n-\ell_1}\ldots \left(\tfrac{R_p}{p!}\right)^{\ell_{p-1}-\ell_p}\left(\tfrac{R_{p+1}}{(p+1)!}\right)^{\ell_p}}{(n-\ell_1)!\ldots (\ell_{p-1}-\ell_p)!\ell_p!}\nonumber\\
&-4\chi\sum_{n=0}^p\frac{(-1)^n(n+1)(p+1-n)}{\hr^{n+2}}\tp{p+1-n}-2\chi\sum_{n=1}^p\frac{(-1)^n(p+2-n)(p+1-n)}{\hr^{n+1}}\,\tp{p+2-n}\bigg]\ .
\end{align}
Together with (\ref{tpCoefsForm}) this solution allows to express the coefficients $\tp{n}$ in the expansion of the deformation function (\ref{ExpansionDeformationFunction}) in terms of $R_0$, $R_1$ and $\{\wtq{n}\}_{n\in \mathbb{N}^*}$. For the $\tau_{0,\ldots,4}$ we obtain
\begin{align}
&\tp{0}=\wtq{0}\,,\hspace{1cm}\tp{1}=R_1\,\wtq{1}\,,\hspace{1cm}\tp{2}=\frac{R_0 \zh}{4\ma}-\frac{R_1}{\zh}\,\wtq{1}\,,\hspace{1cm}\tp{3}=\frac{R_1}{\zh^2}\,\wtq{1}-\frac{R_0-\zh R_1}{12\ma}\,,\nonumber\\
&\tp{4}=\frac{R_0}{96 \zh \ma^2\wtq{1}}\,(\zh^3+12\ma \wtq{1})-R_1\left(\frac{1}{24\ma}+\frac{\wtq{1}}{\zh^3}\right)-\frac{R_1^2\zh \wtq{2}}{48\ma \wtq{1}}\,.\label{TauCoefDistance}
\end{align}
The form of $\tp{4}$ suggests that the coefficients $\tp{n}$ for $n> 4$ are more complicated, however, we remark that (\ref{tpCoefsForm}) implies that they depend linearly on $\wtq{n-2}$ in the following fashion
\begin{align}
&\tp{n}=-\frac{R_1^{n-2} \zh}{2\ma n!}\,\frac{\wtq{n-2}}{\wtq{1}}+\tau^{\text{sub-} \widetilde{R}}_n(R_0,R_1,\wtq{0},\ldots,\wtq{n-3})\,,&&\forall n\geq 4\,,\label{TauExpR}
\end{align}
where $\tau^{\text{sub-} \widetilde{R}}_n$ is independent of $\wtq{k\geq n-2}$.
\subsubsection{\scheme\, identification}
To relate the (redefined) Ricci scalar- and distance \scheme, we next identify the deformation functions $v^{(\rho)}$ in (\ref{SeriesDistanceRho}) with $v^{(\widetilde{R})}$ in (\ref{SeriesExpansionRtzh})
\begin{align}
v^{(\rho)}=\sum_{p=0}^\infty \tp{p}\,(\co-\zh)^p=v^{(\widetilde{R})}\,, \label{SchemeIdentificationRrho}
\end{align}
To this end we express the coefficients $\tp{p}$ in terms of $\{\xi_{2n}\}_{n\in\mathbb{N}^*}$ using the result of~\cite{DelPiano:2023fiw} and taking into account (\ref{ABrestriction})
\begin{align}
&\tp{0}=\frac{\xi_0}{2\ma}\,,\hspace{1cm}\tp{1}=\frac{\xi_2 b_1^2}{2\ma}\,,\nonumber\\
&\tp{p}=\frac{1}{2\ma}\sum_{{0\leq \ell_{p-1}\leq \ldots\leq \ell_1\leq 2n\leq 2p}\atop{n+\ell_1+\ell_2+\ldots+\ell_{p-1}=p}}\frac{(2n)!\,\xi_{2n} b_1^{2n-\ell_1}b_3^{\ell_1-\ell_2}\ldots b_{2p-3}^{\ell_{p-2}-\ell_{p-1}}b_{2p-1}^{\ell_{p-1}}}{(2n-\ell_1)!(\ell_1-\ell_2)!\ldots (\ell_{p-2}-\ell_{p-1})!\ell_{p-1}!}\hspace{1cm}\forall p\geq 2\,.\label{TpDistance}
\end{align}
Using (\ref{Coeffsb}) to express the coefficients $\{b_{2n-1}\}_{n\in\mathbb{N}}$ in terms of the $\{a_{2n-2}\}_{n\in\mathbb{N}}$ and (\ref{Coeffsa}) to express the latter in terms of $\{\xi_{2n-2}\}_{n\in\mathbb{N}}$, we find for the coefficients $\tp{0,\ldots,3}$ of (\ref{TpDistance})
\begin{align}
&\tp{0}=\frac{\xi_0}{2\ma}\,,\hspace{1cm} \tp{1}=\frac{4\zh \xi_2}{(1+\varpi)\ma}\,,\hspace{1cm}\tp{2}=\frac{2\xi_2}{\ma(1+2\varpi)}+\frac{32\zh^2(1+\varpi(3+2\varpi)+8\zh\xi_2)\xi_4}{\ma(1+\varpi)^3(1+2\varpi)}\,,\nonumber\\
&\tp{3}=\frac{640\zh^3 \xi_6}{(1+\varpi)^2(2+3\varpi)\ma}+\frac{2048\zh^4(7+8\varpi) \xi_4^2}{(1+\varpi)^3(1+2\varpi)(2+3\varpi)\ma}+\frac{24\zh(5+9\varpi+6\varpi^2)\xi_4}{(1+\varpi)(1+2\varpi)^2(2+3\varpi)\ma}\nonumber\\
&\hspace{1cm}+\frac{(1+\varpi)^2(1-3\varpi+2\varpi^2)}{16\zh^2(1+2\varpi)^2(2+3\varpi)\ma}\,.\label{TauCoefRicci}
\end{align}
The general form (\ref{ConditionExpansionBs}) of the coefficients $b_{2n-1}$ implies
\begin{align}
\tp{p}=\frac{2^{3p} \zh^p }{2\ma(1+\varpi)^p}\left(\frac{8\xi_2 \zh}{(1+\varpi)((p-1)+p\varpi)}+1\right)\,\xi_{2p}+\tau^{\text{sub-}\rho}_{p}\left(\xi_0,\ldots,\xi_{2p-2}\right)\,,\label{TauExpd}
\end{align}
where $\tau^{\text{sub-}\rho}_{p}$ is independent of $\xi_{2k}$ for $k\geq p+1$. Identifying (\ref{TauExpR}) with (\ref{TauExpd}) allows to relate the coefficients $\{\xi_{2n}\}_{n\in\mathbb{N}^*}$ to $(R_0,R_1,\{\wtq{n}\}_{n\in\mathbb{N}^*})$. Concretely, identifying $\tau_{0,\ldots,3}$ gives relations between $(\xi_0,\xi_2,\xi_4,\xi_6)$ and $(R_0,R_1,\wtq{0},\wtq{1})$, {{which are explicitly given in Appendix~\ref{App:SolutionOfEquationDistRicci}.}} Once these parameters are fixed, the identification of $\tp{p\geq 4}$ (\emph{i.e.} (\ref{TauExpR}) and (\ref{TauExpd})) is a linear relation relating $\xi_{2p}$ and $\wtq{p-2}$, which therefore allows to completely express the distance \scheme\,in terms of the Ricci-scalar \scheme\,and vice versa. Notice that this identification also allows to write physical quantities unambiguously in either \scheme. For example, the Hawking temperature in the distance scheme (\ref{HawkingTempFirst}) can be given more generally in terms of $\tp{0,1}$ as follows
\begin{align}
T_H=\frac{1-2\ma\, \tp{1}}{8\pi \ma\, \tp{0}}\,.\label{HawkingTempGen}
\end{align}
Using (\ref{cp coeff}) and (\ref{TauCoefDistance}) we can therefore write the Hawking temperature in terms of {{the free parameters}} of the Ricci-scalar \scheme
\begin{align}
T_H=\frac{1-2\chi R_1\,\sum_{n=1}^\infty\frac{q_n\,R_0^{n-1}}{(n-1)!}}{8\pi \ma\sum_{m=0}^\infty \frac{q_m R_0^m}{m!}}=\frac{1-2\ma R_1 \wtq{1}}{8\pi\ma \wtq{0}}\,.
\end{align}
Before closing this Section, we briefly remark that the condition $\xi_2\leq \frac{1}{16\zh}$ for the coefficient $\xi_2$ in the distance \scheme\,is automatically satisfied in the Ricci scalar \scheme. Indeed, for any value of $\wtq{1}\,R_1\in\mathbb{R}$, $\xi_2\leq\frac{1}{16\zh}$ is automatically satisfied in (\ref{RicciSchemeInverse}).


\section{Kretschmann Scalar \scheme}\label{Sect:KretschmannScalar}
\subsection{Definition of the \scheme}
Following the example of the Ricci scalar, we can define further \scheme s based on other curvature scalars. As an example of a second-order curvature invariant, we shall consider the Kretschmann scalar \cite{Kretschmann}, which in terms of the Riemann tensor ${R^\mu}_{\nu\rho\lambda}$ is defined as $K=R_{\mu\nu\rho\lambda}R^{\mu\nu\rho\lambda}$, and define the deformed metric function
\begin{align}
f(z)=1-\frac{2\ma}{z}\,v^{(K)}(K(z))\,.
\end{align}
The Kretschmann scalar is related to the deformation function through the following differential equation
\begin{align}
K(z)&=\frac{1}{\co^4}\left(4-8f+4f^2+4\co^2 (f')^2+\co^4 (f'')^2\right)\nonumber\\
&=\frac{8\ma^2}{\co^2} \left[\frac{\dd^2}{\dd\co^2}\left(\frac{(v^{(K)})^2}{\co^2}\right)+\frac{\co^4}{2}\left(\frac{\dd}{\dd \co}\left(\frac{1}{\co^2}\frac{\dd v^{(K)}}{\dd\co}\right)\right)^2\right]\,.\label{SelfConsistencyKretschmann}
\end{align}
Following the example of (\ref{SeriesExpansionszh}), in order to solve this self-consistency relation, we expand the function $v^{(K)}$ as 
\begin{align}
&v^{(K)}(K)=\sum_{n=0}^\infty \frac{\si{n}}{n!}\, K^n\,,&&\text{and} &&K(z)=\sum_{n=0}^\infty \frac{\kc{n}}{n!}\,(\co-\zh)^n\,,\label{KretschmannDofs}
\end{align}
where the coefficients $\{\si{n}\}_{n\in\mathbb{N}^*}$ are external input parameters that {{define}} the \scheme. We choose them in such a way to satisfy the normalisation condition $\sum_{n=0}^\infty \frac{\si{n}}{n!} \kc{0}^n =\frac{\hr}{2\ma}$. As we shall explain, the coefficients $\{K_n\}_{n\in\mathbb{N}^*}$ are a priori not all independent, but related to the $\{\si{n}\}_{n\in\mathbb{N}^*}$ via (\ref{SelfConsistencyKretschmann}). To study these dependencies, we expand $v^{(K)}$ in the Schwarzschild coordinate\footnote{In view of later identifying this \scheme\,with the ones discussed previously, we use the same coefficients $\{\tp{n}\}_{n\in\mathbb{N}^*}$ as in (\ref{v exp r}) and (\ref{SchemeIdentificationRrho}).}
\begin{equation}\label{sigma series sp}
v^{(K)}(K(\co)) = \sum_{p=0}^\infty \tp{p} (\co- \hr)^p \ ,
\end{equation}
where the coefficients $\tp{n}$ can be expressed in terms of the $\{\kc{n}\}_{n\in\mathbb{N}^*}$ as follows
\begin{align}
\tp{0}&=\sum_{n=0}^\infty \frac{\si{n}}{n!}\,(K_0)^n\,,\nonumber\\
 \tp{p}&= \sum_{n=0}^{\infty} \si{n} \sum_{{0\leq \ell_p\leq \ell_{p-1}\leq\ldots\leq \ell_1\leq n}\atop{\ell_1+\ell_2+\ldots+\ell_p=p}}\frac{\left(\kc{0}\right)^{n-\ell_1}\,\left(\kc{1}\right)^{\ell_1-\ell_2}\ldots\left(\frac{\kc{p}}{p!}\right)^{\ell_p}}{(n-\ell_1)!\,(\ell_1-\ell_2)!\ldots (\ell_{p-1}-\ell_p)!\ell_p!}\,.\label{coeff s}
\end{align}
Furthermore, from (\ref{SelfConsistencyKretschmann}) we obtain
{\allowdisplaybreaks
\begin{align}
&\frac{\kc{p}}{p!}=8\ma^2\,\sum_{m=0}^p\sum_{n=0}^{m+2}\sum_{\ell=0}^{m+2-n}\frac{(-1)^{p+\ell-m}(p-m+1)(\ell+1)(m+2)(m+1)}{\zh^{p-m+\ell+4}}\,\tp{n}\,\tp{m+2-n-\ell}\nonumber\\
&+4\ma^2\sum_{n=1}^{p-1}\sum_{\ell=0}^{p-n}\sum_{k=0}^n\frac{(-1)^{\ell+k}(\ell+1)(k+1)(p-n)n(p-n+1-\ell)(n+1-k)}{\zh^{k+\ell+4}}\,\tp{p-n-\ell+1}\,\tp{n-k+1}\nonumber\\
&+8\ma^2\sum_{n=1}^{p}\sum_{\ell=0}^{p-n+1}\sum_{k=0}^n\frac{(-1)^{\ell+k}(\ell+1)(k+1)(p-n+1)n(p-n+2-\ell)(n+1-k)}{\zh^{k+\ell+3}}\,\tp{p-n-\ell+2}\,\tp{n-k+1}\nonumber\\
&+4\ma^2\sum_{n=1}^{p+1}\sum_{\ell=0}^{p-n+2}\sum_{k=0}^n\frac{(-1)^{\ell+k}(\ell+1)(k+1)(p-n+2)n(p-n+3-\ell)(n+1-k)}{\zh^{k+\ell+2}}\,\tp{p-n-\ell+3}\,\tp{n-k+1}\,.\label{FullRecursionKretschmann}
\end{align}}
Notice that for $p>0$ 
\begin{align}
&\frac{\kc{p}}{p!}=\frac{16\ma^2(p+1)(p+2)}{\zh^4}\,\left(\tp{0}-\zh\,\tp{1}+\zh^2\,\tp{2}\right)\,\tp{p+2}+\widehat{\tau}^{\text{sub-}K}_{p+2}(\tp{0},\ldots,\tp{p+1})\nonumber\\
&=\frac{16\ma^2 \kc{p+2}}{\zh^4 p!}\,\left(\sum_{m=0}^\infty\si{m} \kc{0}^{m-2}\left(\frac{\kc{0}^2}{m!}-\frac{\zh \kc{0} \kc{1}}{(m-1)!}+\frac{\zh^2 \kc{0} \kc{2}}{2(m-1)!}+\frac{\zh^2 \kc{1}^2}{2(m-2)!}\right)\right)\,\left(\sum_{n=1}^\infty \frac{\si{n} (\kc{0})^{n-1}}{(n-1)!}\right)\nonumber\\
&\hspace{3cm}+\tau_{p+2}^{\text{sub-}K}(\kc{0},\ldots,\kc{p+1},\{\si{n}\})\,,\label{KretschmannRecursion}
\end{align}
where $\widehat{\tau}^{\text{sub-}K}_{p+2}$ and $\tau_{p+2}^{\text{sub-}K}$ denote terms independent of $\tp{n>p+2}$ and $\kc{n>p+2}$ respectively. Furthermore, in the second line we have used an expansion of the coefficients $\tp{p+2}$ similar to (\ref{Leading1}) or (\ref{TauExpR}). We therefore find that (\ref{KretschmannRecursion}) is a linear equation for $\kc{p+2}$, which, similar to the case of the Ricci scalar \scheme, allows to express the $\kc{p+2}$ uniquely in terms of $\kc{p'<p+2}$ for all $p\geq 3$. For $p=0$, (\ref{FullRecursionKretschmann}) is a quadratic equation for $\kc{2}$ with the following solutions 
\begin{align}
\kc{2}=\frac{2\zeta_1(\kc{1}(2\zeta_1-\kc{1}\zh \zeta_2)\ma-1)+u|\zeta_1|\sqrt{16 \kc{1}\zeta_1\chi(1-\kc{1}\zeta_1\ma)-8+\kc{0}\zh^4}}{2\ma\zh \zeta_1^2}\,,\label{K2solt}
\end{align}
where we have introduced a phase factor $u=\pm1$ and defined the following shorthand nota-
\begin{wrapfigure}{r}{0.32\textwidth}
${}$\\[-1cm]
\begin{center}
\scalebox{1}{\parbox{5cm}{\begin{tikzpicture}        
\draw[rounded corners,fill=yellow!45!white] (-2.5, -0.1) rectangle (2.5, -3.1);
\node at (0,-0.5) {\bf Kretschmann scalar $K$};
\node at (-0.45,-1) {\tiny param. (\ref{KretschmannDofs}): $\kc{0}\,,\kc{1}\,,\{\si{n}\}_{n\in\mathbb{N}^*}$};
\node at (-0.6,-1.4) {\tiny self-consistency equation (\ref{SelfConsistencyKretschmann}):};
\node at (0,-1.9) {\tiny $K=\frac{\left(4-8f+4f^2+4\co^2 (f')^2+\co^4 (f'')^2\right)}{\co^4}$};
\node at (-0.95,-2.4) {\tiny  normalisation condition:};
\node at (0,-2.8) {\tiny $K(\zh)=\kc{0}$, $u=\pm1$};
\end{tikzpicture}
}}
\caption{Summary of the Kretschmann scalar \scheme.}
\label{Fig:SummaryKretschmann}
\end{center}
${}$\\[-2.5cm]
\end{wrapfigure}
\noindent
tion
\begin{align}
&\zeta_1=\sum_{n=1}^\infty \si{n}\frac{\kc{0}^{n-1}}{(n-1)!}\,,\nonumber\\
&\zeta_2=\sum_{n=2}^\infty \si{n}\frac{\kc{0}^{n-2}}{(n-2)!}\,.\label{ZetaDefs}
\end{align}
Furthermore, we have used $\tp{0}=\frac{\zh}{2\ma}$ to simplify the expression. Based on this solution, the Kretschmann scalar \scheme\,is summarised in Figure~\ref{Fig:SummaryKretschmann}. Here we consider as {{free parameters}} $(\kc{0},\kc{1},\{\si{n}\}_{n\in\mathbb{N}^*})$ (which are real parameters), while the phase factor $u=\pm1$ is an additional condition. This is similar to our choice for dealing with a sign ambiguity when resolving the self-consistency condition in the distance \scheme. Finally, we remark that with (\ref{coeff s}) the Hawking temperature (\ref{HawkingTempGen}) in the Kretschmann-scalar \scheme\,takes the following form
\begin{align}
T_H=\frac{1-2\chi \kc{1}\,\sum_{n=1}^\infty\frac{\si{n}\,\kc{0}^{n-1}}{(n-1)!}}{8\pi \ma\sum_{m=0}^\infty \frac{\si{m} \kc{0}^m}{m!}}=\frac{1-2\chi \kc{1}\,\zeta_1}{4\pi \zh}\,.
\end{align}

\subsection{Relation to Other \scheme s}
After having introduced the Kretschmann scalar \scheme, we now relate it to other \scheme s. To this end, following the example of the Ricci scalar \scheme, we shall first propose a slight redefinition, before making contact to the {{distance- and}} Ricci scalar \scheme.

Instead of working with the Kretschmann scalar directly, we instead subtract from it its value at the horizon, \emph{i.e.} we introduce $\widetilde{K}(z):=K(z)-\kc{0}$ and define
\begin{align}
&v^{(\widetilde{K})}(\widetilde{K})=\sum_{n=0}^\infty \frac{\sit{n}}{n!}\, \widetilde{K}^n\,,&&\text{and} &&\widetilde{K}(z)=\sum_{n=1}^\infty \frac{\kc{n}}{n!}\,(\co-\zh)^n\,,
\end{align}
The relation between the coefficients $\{\si{n}\}_{n\in\mathbb{N}^*}$ and $\{\sit{n}\}_{n\in\mathbb{N}^*}$ follows in the same manner as in (\ref{SchemeReductionR}), namely
\begin{align}
&\si{p}=\sum_{n=p}^\infty\frac{\sit{n}\,(-\kc{0})^{n-p}}{(n-p)!}\,,&&\text{and}&&\sit{p}=\sum_{n=p}^\infty\frac{\si{n}\,\kc{0}^{n-p}}{(n-p)!}\,.\label{SchemeReductionK}
\end{align}
To make contact with other \scheme s, we expand $v^{(\widetilde{K})}$ in terms of the Schwarzschild coordinate
\begin{align}
v^{(\widetilde{K})}(\widetilde{R})=\sum_{p=0}^\infty \tp{p}\,(\co-\zh)^p\,. \label{ExpansionDeformationFunctionKretschmann}
\end{align}
and interpret the coefficients $\tp{p}$ to be the same as in (\ref{v exp r}), (\ref{SchemeIdentificationRrho}) and (\ref{sigma series sp}), while being expanded as a function of the $\{\sit{n}\}_{n\in\mathbb{N}^*}$ and $\{\kc{n}\}_{n\in\mathbb{N}}$ 
\begin{align}
\tp{0}&=\sit{0}\,,\nonumber\\
\tp{p}&=\sum_{{0\leq \ell_{p-1}\leq\ldots\leq \ell_1\leq n\leq p}\atop{\ell_1+\ell_2+\ldots+\ell_{p-1}+n=p}} \frac{\sit{n}\,\left(\tfrac{\kc{1}}{1!}\right)^{n-\ell_1}\left(\tfrac{\kc{2}}{2!}\right)^{\ell_1-\ell_2}\ldots \left(\tfrac{\kc{p-1}}{(p-1)!}\right)^{\ell_{p-2}-\ell_{p-1}}\left(\tfrac{\kc{p}}{p!}\right)^{\ell_{p-1}}}{(n-\ell_1)!\,(\ell_1-\ell_2)!\ldots (\ell_{p-2}-\ell_{p-1})!\ell_{p-1}!}\hspace{0.5cm} \forall p\geq 1\,.\label{tpCoefsFormKretschmann}
\end{align}
Using (\ref{FullRecursionKretschmann}) we can explicitly express the $\tp{p}$ in terms of $\{\sit{n}\}_{n\in\mathbb{N}^*}$ and $(\kc{0},\kc{1},u)$, \emph{e.g.} we find for the first few instances
\begin{align}
&\tp{0}=\sit{0}\,,\hspace{1cm}\tp{1}=\kc{1}\,\sit{1}\,,\hspace{1cm}\tp{2}=\frac{4\ma \sit{1}(\kc{1}\zh \sit{1}-\sit{0})+\mathfrak{p}}{4\ma\sit{1}\zh}\,,\nonumber\\
&\tp{3}=\frac{6 \kc{0} \sit{1} \zh ^6-64 \kc{1}^2 \sit{1}^3 \ma^2 \zh ^2+\kc{1} \sit{1} \zh  \left(-16
   \mathfrak{p} \ma+96 \sit{0} \sit{1} \ma^2+\zh ^6\right)+16 \sit{0} \ma (\mathfrak{p}-4 \sit{0}
   \sit{1} \ma)}{24 \mathfrak{p} \ma \zh ^3}\,,\label{TpKretschLow}
\end{align}
with the abbreviation $\mathfrak{p}=u |\sit{1}|\sqrt{\kc{0}\zh^6-16(2\sit{0}^2-2\kc{1}\zh\sit{0}\sit{1}+\kc{1}^2\zh^2\sit{1}^2)\ma^2}$. Higher orders in $\tp{n}$ are more complicated, however, we remark that they depend linearly on $\sit{n-2}$
\begin{align}
&\tp{n}=-\frac{\kc{1}^{n-2} \zh^4}{2\ma n! 2\mathfrak{p}}\,\sit{n-2}+\tau^{\text{sub-} \widetilde{K}}_n(\kc{0},\kc{1},u,\sit{0},\ldots,\sit{n-3})\,,&&\forall n\geq 4\,,\label{TauExpKtilde}
\end{align}
where $\tau^{\text{sub-} \widetilde{K}}_n$ is independent of $\{\sit{k\geq n-1}\}$. Identifying the $\tp{p}$ in (\ref{TpKretschLow}) and (\ref{TauExpKtilde}) with (\ref{v exp r}), (\ref{SchemeIdentificationRrho}) and (\ref{sigma series sp}) yields equations that allow to relate the Kretschmann scalar \scheme\,to the distance- and Ricci scalar \scheme. Indeed, identifying $\tp{0,1,2,3}$ fixes $(\kc{0},\kc{1},\sit{0},\sit{1})$ in terms of $(R_0,R_1,\wtq{0},\wtq{1})$ or $(\xi_0,\xi_2,\xi_4,\xi_6)$ (and vice-versa). Notice, that the space of solutions for these equations is fairly restricted: for example, for given $(\tp{0},\tp{1},\tp{2},\tp{3})$ (either given in terms of $(\xi_0,\xi_2,\xi_4,\xi_6)$ or $(R_0,R_1,\wtq{0},{\wtq{1}})$), consistent solutions of $(\kc{0},\kc{1},\sit{0},\sit{1})$ not only require a specific choice of $u$, but they are in fact uniquely fixed. To see this, we realise that the third equation in (\ref{TpKretschLow}) can be rewritten in the form
\begin{align}
\frac{u\,\text{sign}(\sit{1})}{4\chi\zh}\,\sqrt{\kc{0}\zh^6-16(2\tp{0}^2-2\tp{1}\tp{0}\zh+\tp{1}^2\zh^2)\ma^2}=\tp{2}-\tp{1}+\frac{\tp{0}}{\zh}\,,\label{RecSignDem}
\end{align}
which only has a solution if $u\, \text{sign}(\sit{1})=\text{sign}\left(\tp{2}-\tp{1}+1\right)$, in which case $K_0$ is uniquely fixed. This, however, renders the last equation in (\ref{TpKretschLow}) linear in $\sit{1}$, with the solution
\begin{align}
\sit{1}=\frac{\zh^6 \tp{1}}{32\zh(\tp{1}-2\tp{2}+2\tp{1}\tp{2}+3\zh^2\tp{3}(1-\tp{1}+\tp{2})-6K_0}\,.
\end{align}
Since this fixes (the sign of) $\sit{1}$, (\ref{RecSignDem}) is consistent only for a specific value of $u$. Finally, as in the previous cases, once (\ref{TpKretschLow}) is solved, (\ref{TauExpKtilde}) are linear equations for $\sit{n-2}$ in terms of $\tp{n}$. Due to the length of the ensuing expressions, we refrain from reproducing the explicit form of the solution in this paper\footnote{For the same reason, we also do not provide the (unique) inverse transformations that express $(\xi_0,\xi_2,\xi_4,\xi_6)$ or $(R_0,R_1,\wtq{0},\wtq{1})$ in terms of $(K_0,K_1,\sit{0},\sit{1})$)): they follow from replacing $\tp{n}$ in (\ref{TauCoefRicci}), (\ref{TauExpd}) and (\ref{TauCoefDistance}), (\ref{TauExpR}) respectively with (\ref{TpKretschLow}) and (\ref{TauExpKtilde}). We furthermore remark in passing that, similar to the case of the Ricci scalar \scheme, the condition $\xi_2\leq \frac{1}{16\zh}$ also poses no further conditions on $(K_0,K_1,\{\sit{n}\}_{n\in\mathbb{N}^*})$.}, however, we provide a concrete mapping from the distance to the Kretschmann scalar \scheme\,for a particular small deformation of the Schwarzschild solution in Appendix~\ref{App:KretschmannSigns}. Higher orders $\tp{n\geq 4}$ are linear in $\sit{n-2}$ and thus uniquely allow to fix them in terms of $\wtq{n-2}$ or $\xi_{2n}$ (or vice-versa).

\section{Summary of Results and Conclusions}\label{Sect:Conclusions}
In this paper we have defined \scheme s as coherent frameworks to describe (quantum) deformed {{-- relative to the classical Schwarzschild geometry \cite{Schwarzschild} -- }} spherically symmetric and static black holes in 4 dimensions. Following previous work \cite{Binetti:2022xdi,DelPiano:2023fiw}, such black hole geometries are captured by one\footnote{In this work we only focused on the case of a single deformation function. The more general case of two such functions, which is compatible with spherical symmetry and static behaviour of the black hole, can be tackled using the same methods described here.} deformation function of a physical quantity $\mathcal{X}$ to ensure compatibility of the quantum geometry with the same symmetries as its classical counterpart. 

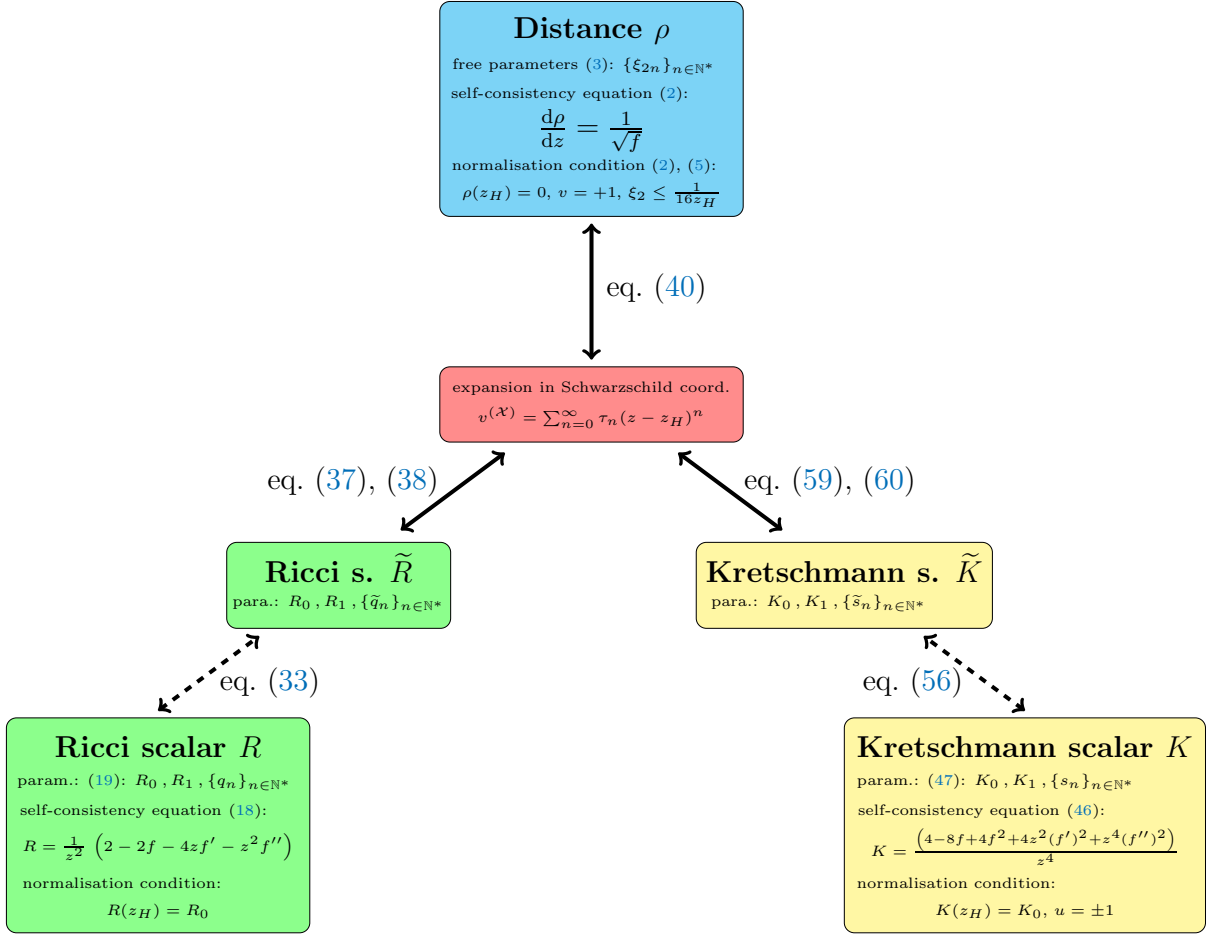
\begin{figure}[t]
\begin{center}
\scalebox{0.95}{\parbox{16.6cm}{\begin{tikzpicture}        
\draw[rounded corners,fill=cyan!45!white] (-2.1, -0.1) rectangle (2.1, -3.1);
\node at (0,-0.5) {\bf Distance $\rho$};
\node at (-0.1,-1) {\tiny free parameters (\ref{SeriesDistanceRho}): $\{\xi_{2n}\}_{n\in\mathbb{N}^*}$};
\node at (-0.3,-1.4) {\tiny self-consistency equation (\ref{SelfEquationDistance}):};
\node at (0,-1.9) {$\frac{\dd\rho}{\dd\co}=\frac{1}{\sqrt{f}}$};
\node at (-0.1,-2.4) {\tiny  normalisation condition (\ref{SelfEquationDistance}), (\ref{Coeffsa}):};
\node at (0,-2.8) {\tiny $\rho(\zh)=0$, $v=+1$, $\xi_2\leq\frac{1}{16\zh}$};
\begin{scope}[yshift=-5cm]
\draw[rounded corners,fill=red!45!white] (-2.1, -0.2) rectangle (2.1, -1.25);
\node at (0,-0.5) {\tiny expansion in Schwarzschild coord.};
\node at (0,-0.9) {\tiny $v^{(\mathcal{X})}=\sum_{n=0}^\infty \tp{n}(\co-\zh)^n$};
\end{scope}
\begin{scope}[xshift=-6cm,yshift=-10cm]
\draw[rounded corners,fill=green!45!white] (-2.1, -0.1) rectangle (2.1, -3.1);
\node at (0,-0.5) {\bf Ricci scalar $R$};
\node at (-0.05,-1) {\tiny param.: (\ref{SeriesExpansionszh}): $R_0\,,R_1\,,\{q_n\}_{n\in\mathbb{N}^*}$};
\node at (-0.2,-1.4) {\tiny self-consistency equation (\ref{FormRicciEq}):};
\node at (0,-1.9) {\tiny $R=\frac{1}{\co^2}\,\left(2-2f -4\co f'-\co^2 f''\right)$};
\node at (-0.5,-2.4) {\tiny  normalisation condition:};
\node at (0,-2.8) {{\tiny $R(\zh)=R_0$}};
\end{scope}
\begin{scope}[xshift=-3.5cm,yshift=-7.5cm]
\draw[rounded corners,fill=green!45!white] (-1.55, -0.15) rectangle (1.55, -1.35);
\node at (0,-0.5) {\bf Ricci s. $\widetilde{R}$};
\node at (0,-1) {\tiny para.: $R_0\,,R_1\,,\{\wtq{n}\}_{n\in\mathbb{N}^*}$};
\end{scope}
\begin{scope}[xshift=3.5cm,yshift=-7.5cm]
\draw[rounded corners,fill=yellow!45!white] (-2.05, -0.15) rectangle (2.05, -1.35);
\node at (0,-0.5) {\bf Kretschmann s. $\widetilde{K}$};
\node at (-0.35,-1) {\tiny para.: $\kc{0}\,,\kc{1}\,,\{\sit{n}\}_{n\in\mathbb{N}^*}$};
\end{scope}
\begin{scope}[xshift=6cm,yshift=-10cm]    
\draw[rounded corners,fill=yellow!45!white] (-2.5, -0.1) rectangle (2.5, -3.1);
\node at (0,-0.5) {\bf Kretschmann scalar $K$};
\node at (-0.4,-1) {\tiny param.: (\ref{KretschmannDofs}): $\kc{0}\,,\kc{1}\,,\{\si{n}\}_{n\in\mathbb{N}^*}$};
\node at (-0.6,-1.4) {\tiny self-consistency equation (\ref{SelfConsistencyKretschmann}):};
\node at (0,-1.9) {\tiny $K=\frac{\left(4-8f+4f^2+4\co^2 (f')^2+\co^4 (f'')^2\right)}{\co^4}$};
\node at (-0.95,-2.4) {\tiny  normalisation condition:};
\node at (0,-2.8) {\tiny $K(\zh)=\kc{0}$, $u=\pm1$};
\end{scope}
\draw[ultra thick,<->] (0,-3.2) -- (0,-5.1);
\node at (0.9,-4.1) {eq.~(\ref{TpDistance})};
\draw[ultra thick,<->] (-1.2,-6.4) -- (-2.64,-7.48);
\node at (-3.3,-6.8) {eq.~(\ref{TauCoefDistance}), (\ref{TauExpR})};
\draw[ultra thick,<->,dashed] (-6,-10) -- (-4.6,-8.96);
\node at (-4.45,-9.6) {eq.~(\ref{SchemeReductionR})};
\draw[ultra thick,<->] (1.2,-6.4) -- (2.64,-7.48);
\node at (3.3,-6.8) {eq.~(\ref{TpKretschLow}), (\ref{TauExpKtilde})};
\draw[ultra thick,<->,dashed] (6,-10) -- (4.6,-8.96);
\node at (4.45,-9.6) {eq.~(\ref{SchemeReductionK})};
\end{tikzpicture}
}}
\end{center}
\caption{Relation between the three \scheme s discussed in this work.}
\label{Fig:Overview}
\end{figure}

In \cite{Binetti:2022xdi,DelPiano:2023fiw} (inspired by \cite{Bonanno:2000ep}) the physical distance, measured from the (horizon of the) black hole, was chosen. However, other physical quantities (\emph{e.g.} curvature scalars of different orders) are equally viable. In each case, in order to remain self-consistent, $\mathcal{X}$ is computed from the deformed metric, thus leading to an involved equation to tackle (see eqs.~(\ref{SelfEquationDistance}), (\ref{FormRicciEq}) and (\ref{SelfConsistencyKretschmann}) for the distance, Ricci scalar and Kretschmann scalar respectively). To solve this equation, we follow the general framework developed in \cite{DelPiano:2023fiw} and consider series expansions in the vicinity of the event horizon of the black hole. In this way, we not only provide self-consistent solutions for metric deformations based on the physical distance, Ricci- and Kretschmann scalar, but also propose clear definitions of each of the \scheme s, as shown in the overview in Figure~\ref{Fig:Overview}. Moreover, as also indicated in the schematic drawing, by expressing the deformation functions as expansions in the Schwarzschild coordinate, we find equations which allow to relate the different \scheme s to one another. In the case of the \scheme s based on curvature invariants, we first perform a slight re-definition of the \scheme\,((\ref{SchemeReductionR}) and (\ref{SchemeReductionK}) respectively), by subtracting the value at the horizon: in this way, the equations relating the (modified) \scheme\, to the distance \scheme, can be solved in an iterative fashion.

From a conceptual perspective, the relation between the distance \scheme\, and \scheme s based on curvature scalars is non-trivial: indeed, the former is a non-local physical quantity (since it is defined as an integral between two points in space-time), while the latter are local invariants. Nevertheless, at least sufficiently close to the horizon, they are both expected to be monotonic functions of the Schwarzschild coordinate, which thus allows the mapping between the \scheme s.

We furthermore also remark that a basic assumption of our work is a sufficient level of regularity of the geometry close to the horizon, such that series expansions remain well-defined. While here we have defined \scheme s based on full (\emph{i.e.} infinite) series expansions with suitable radius of convergence, in practice the relations among different \scheme s remain (approximately) valid if these series are truncated after certain orders. This property is useful for practical computations, when geometries are approximated by their leading coefficients (see \emph{e.g.} \cite{DAlise:2023hls} for an example). It would be interesting to generalise our framework beyond the level of series expansions: indeed, in this way \scheme s (and their relations) could be extended to larger regions of the space-time (not restricted by the radius of convergence). Similarly, in the current work (following similar assumptions as in \cite{DelPiano:2023fiw}), we have restricted ourselves to the region outside of the (simple) horizon of the black hole. Generalising our methods to the interior of the black hole will allow us to study interesting new physical questions (such as the fate of the central singularity in the quantum theory) from a different perspective.

Our work can be generalised to incorporate more sophisticated geometries (\emph{e.g.} charged- \cite{reissner_2018_1447315,weyl_2018_1424330,Nordstrom} or rotating black holes \cite{Kerr}): indeed, {{using different \scheme s opens the possibility to describe quantum deformed black holes in new ways that may be better adapted to capture the particularities of specific geometries.}} 
\section*{Acknowledgements}
{{We would like to thank Emmanuele Battista,  Mattia Damia Paciarini, Gerald Dunne, Aaron Held  and Mikolaj Myszkowski  for useful discussions on related topics. 
The work of F.S. is
partially supported by the Carlsberg Foundation, semper ardens grant CF22-0922.
}}

\vskip 2cm

\appendix

\section{Power Series Expansion}\label{App:PowerPower}
In this appendix we illustrate how to extract a specific term from a(n integer) power of a series. Indeed, let
\begin{align}
&S(x)=\sum_{k=0}^\infty c_k\,x^k\,,&&\forall |x|\leq x_c 
\end{align}
where $x_c$ is the radius convergence, $c_k\in\mathbb{R}$ a set of expansion coefficients and let $n\in\mathbb{N}$. We are then interested in the expansion 
\begin{align}
(S(x))^n=\sum_{k=0}^\infty \mathfrak{c}_k\,x^k\,,\label{PowerPowerExpansion}
\end{align}
where the $\mathfrak{c}_k$ depend explicitly on $\{c_k\}_{k\in\mathbb{N}^*}$. To calculate $\mathfrak{c}_p$ (for given $p\in\mathbb{N}$), we consider
\begin{align}
(S(x))^n=\left(\sum_{k=0}^p c_k\,x^k\right)^n+\mathcal{O}(x^{p+1})=\sum_{\ell_0=0}^n\, \left(\begin{linesmall}{@{}c@{}} \\[-6pt]n \\ \ell_0 \end{linesmall}\right)\,c_0^{n-\ell_0}\,x^{\ell_0}\,\left(\sum_{k=1}^p c_k x^{k-1}\right)^{\ell_0}\,,
\end{align}
where we have used the binomial formula. This expression can be further re-written
\begin{align}
(S(x))^n=\sum_{\ell_0=0}^n\, \left(\begin{linesmall}{@{}c@{}} \\[-6pt]n \\ \ell_0 \end{linesmall}\right)\,c_0^{n-\ell_0}\,x^{\ell_0}\sum_{\ell_1=0}^{\ell_0}\,\left(\begin{linesmall}{@{}c@{}} \\[-6pt]\ell_0 \\ \ell_1 \end{linesmall}\right)\,c_1^{\ell_0-\ell_1}\,x^{\ell_1}\,\left(\sum_{k=2}^p c_k x^{k-2}\right)^{\ell_1}\,.
\end{align}
Iterating this procedure, we find
\begin{align}
(S(x))^n=\sum_{\ell_0=0}^n\sum_{\ell_1=0}^{\ell_0}\ldots\sum_{\ell_{p-1}}\, \left(\begin{linesmall}{@{}c@{}} \\[-6pt]n \\ \ell_0 \end{linesmall}\right)\left(\begin{linesmall}{@{}c@{}} \\[-6pt]\ell_0 \\ \ell_1 \end{linesmall}\right)\ldots \left(\begin{linesmall}{@{}c@{}} \\[-6pt]\ell_{p-2} \\ \ell_{p-1} \end{linesmall}\right)\,x^{\ell_0+\ell_1+\ldots+\ell_{p-1}}\,c_0^{n-\ell_0}\,c_1^{\ell_0-\ell_1}\ldots c_{p-1}^{\ell_{p-2}-\ell_{p-1}}\,c_p^{\ell_{p-1}}\,.
\end{align}
Thus, the coefficient $\mathfrak{c}_p$ in (\ref{PowerPowerExpansion}) is given by
\begin{align}
\mathfrak{c}_0&=c_0^n\,,\nonumber\\
\mathfrak{c}_p&=\sum_{{0\leq \ell_{p-1}\leq \ell_{p-2}\leq \ldots\leq \ell_0\leq n}\atop{\ell_0+\ell_1+\ldots+\ell_{p-1}=p}}\,\frac{n!\,c_0^{n-\ell_0}c_1^{\ell_0-\ell_1}\ldots c_{p-1}^{\ell_{p-2}-\ell_{p-1}}c_p^{\ell_{p-1}}}{(n-\ell_0)!(\ell_0-\ell_1)!\ldots (\ell_{p-2}-\ell_{p-1})!\ell_{p-1}!}\,,&&\forall p\geq 1\,.
\end{align}
For $c_0=0$, the summation over $\ell_0$ is fixed to $\ell_0=n$, such that in this case
\begin{align}
\mathfrak{c}_0&=0\,,\hspace{1cm}\mathfrak{c}_1=c_1\,\delta_{n,1}\,,\nonumber\\
\mathfrak{c}_p&=\sum_{{0\leq \ell_{p-1}\leq \ell_{p-2}\leq \ldots\leq \ell_1\leq n}\atop{\ell_1+\ldots+\ell_{p-1}=p-n}}\,\frac{n!\,c_1^{n-\ell_1}\ldots c_{p-1}^{\ell_{p-2}-\ell_{p-1}}c_p^{\ell_{p-1}}}{(n-\ell_1)!\ldots (\ell_{p-2}-\ell_{p-1})!\ell_{p-1}!}\,,&&\forall p\geq 2\,.\label{DefRedFormPower}
\end{align}

\section{Transition from Ricci to Distance \scheme}\label{App:SolutionOfEquationDistRicci}
Identifying the coefficients $\{\tp{n}\}_{n=0,1,2,3}$ in (\ref{TauCoefDistance}) with those in (\ref{TauCoefRicci}) leads to the following solutions of $(R_0,R_1,\wtq{0},\wtq{1})$ in terms of $(\xi_0,\xi_2,\xi_4,\xi_6)$ 
{\allowdisplaybreaks
\begin{align}
R_0&=\frac{3+5\varpi-3\varpi^2-5\varpi^3+384 \zh^3\xi_4}{2\zh^2(1+\varpi)(1+2\varpi)}\,,\nonumber\\
R_1&=\frac{7680\zh^2}{(1+\varpi)^2(2+3\varpi)}\,\xi_6-\frac{9+61\varpi+73\varpi^2-53\varpi^3-90\varpi^4}{4\zh^3(1+2\varpi)^2(2+3\varpi)}\nonumber\\
&\hspace{0.5cm}+\frac{96(19+41\varpi+30\varpi^2)}{(1+\varpi)(1+2\varpi)^2(2+3\varpi)}\,\xi_4+\frac{24576\zh^3(7+8\varpi)}{(1+\varpi)^3(1+2\varpi)^2(2+3\varpi)}\,\xi_4^2\,,\nonumber\\
\wtq{0}&=\frac{\xi_0}{2\ma}\,,\nonumber\\
\wtq{1}&=\bigg[\frac{30720 \zh^2\ma \xi_6}{(1-\varpi)(1+\varpi)^2(2+3\varpi)\xi_2}-\frac{(9+70\varpi+143\varpi^2+90\varpi^3)\ma}{\zh^3(1+2\varpi)^2(2+3\varpi)}\nonumber\\
&\hspace{0.5cm}+\frac{384(19+41\varpi+30\varpi^2)\ma \xi_4}{(1-\varpi)(1+\varpi)(1+2\varpi)^2(2+3\varpi)}+\frac{98304\zh^3(7+8\varpi)\ma \xi_4^2}{(1-\varpi)(1+\varpi)^3(1+2\varpi)^2(2+3\varpi)}\bigg]^{-1}
\end{align}}
These relations can also be inverted to express $(\xi_0,\xi_2,\xi_4,\xi_6)$ in terms of $(R_0,R_1,\wtq{0},\wtq{1})$ as follows
{\allowdisplaybreaks
\begin{align}
\xi_0&=2\ma\,\wtq{0}\,,\nonumber\\
\xi_2&=\frac{\wtq{1} R_1 \ma}{2\ma}\,(1-2\wtq{1}\,R_1\,\ma)\,,\nonumber\\
\xi_4&=\frac{(1+\sigma)^3\left(R_0\zh^2(1+2\sigma)-8\wtq{1} R_1\ma(1+\sigma-\wtq{1}R_1 \ma)\right)}{384\zh^3\left(1+\sigma-4\wtq{1}R_1\ma+8\wtq{1}R_1^2\ma^2\right)}\,,\nonumber\\
\xi_6&=\frac{(1+\sigma)^4(2+3\sigma)\ma}{46080\zh^3 u}\bigg[\frac{36\wtq{1} R_1}{\zh^2}-\frac{3R_0}{\ma}+\frac{3R_1\zh}{\ma}-\frac{(1+\sigma)^2}{\zh^2\ma(1+2\sigma)u^2}\nonumber\\
&\hspace{3cm}\times \left(R_0\zh^2(1+2\sigma)-8\wtq{1}R_1\ma(1+\sigma+\wtq{1}R_1\ma)\right)\nonumber\\
&\hspace{3cm}\times\left(R_0\zh^2(1+2\sigma)+(1+\sigma-4\wtq{1}R_1\ma)(3-8\wtq{1}R_1\ma)\right)\nonumber\\
&\hspace{1cm}-\frac{2\wtq{1}R_1(1+\sigma)(1-2\wtq{1}R_1\ma)}{u^2 \zh^2(1+2\sigma)^2(2+3\sigma)}\bigg(2 R_0 ^2 \zh^4 \left(32 \wtq{1} R_1  (4 \sigma +7) \ma -32 \sigma
   -31-64 \wtq{1}^2 R_1 ^2 (4 \sigma +7) \ma ^2\right)\nonumber\\
   &\hspace{3cm}+R_0  \zh^2 (\sigma  (16 \wtq{1} R_1  \ma  (2 \wtq{1} R_1  \ma  (128 \wtq{1} R_1  \ma -107)+75)-117)\nonumber\\
   &\hspace{3cm}+4
   \wtq{1} R_1  \ma  (417-2 \wtq{1} R_1  \ma  (64 \wtq{1} R_1  \ma  (17 \wtq{1} R_1  \ma -31)+1013))-117)\nonumber\\
   &\hspace{3cm}-2 (3-8
   \wtq{1} R_1  \ma )^2 (4 \wtq{1} R_1  \ma  (17 \wtq{1} R_1  \ma -7)-1) (-4 \wtq{1} R_1  \ma +\sigma +1)\bigg)\bigg]\label{RicciSchemeInverse}
\end{align}}
with $\sigma=|1-4\wtq{1}R_1 \ma|$ and $u=1+\sigma-4\wtq{1}R_1\ma(1-\wtq{1}R_1\ma)$.
\section{Signs in the Kretschmann Scalar \scheme}\label{App:KretschmannSigns}
The sign choice appearing in the definition of the Kretschmann scalar \scheme\,in Section~\ref{Sect:KretschmannScalar} a priori have an impact on the type of solutions that can be described within this \scheme. Indeed, in Section~\ref{Sect:ReviewDistance}, following \cite{DelPiano:2023fiw}, we have fixed a similar choice in eq.~(\ref{Coeffsa}) by demanding that the corresponding geometry allows a smooth limit recovering the Schwarzschild space-time. A similar condition also fixes certain signs in the Kretschmann scalar \scheme\,(and its connection to the distance \scheme). To illustrate this idea, we start with the distance \scheme\, summarised in Figure~\ref{Fig:SummaryDistance} for the particular choice $\xi_{2n}=0$ $\forall n>1$, such that the only non-trivial remaining parameter is $\xi_2$, or equivalently $\varpi$. The classical limit, which indeed reproduces the Schwarzschild space-time, is $\varpi\to 1$. For this particular choice of parameters we find
\begin{align}
&\tp{0}=\frac{\zh}{2\ma}\,,&&\tp{1}=\frac{1-\varpi^2}{4\ma(1+\varpi)}\,,&&\tp{2}=\frac{1-\varpi^2}{8\ma\zh(1+2\varpi)}\,,&&\tp{3}=\frac{(1-\varpi)(1+\varpi)^2(1-2\varpi)}{16\ma\zh^2(2+3\varpi)(1+2\varpi)^2}\,.\nonumber
\end{align}
Thus, identification with the Kretschmann scalar \scheme\,leads to
\begin{align}
K_0&=\frac{29+124\varpi+170\varpi^2+84\varpi^3+25\varpi^4}{4\zh^4(1+2\varpi)^2}=\frac{12}{\zh^4}+\mathcal{O}(\varpi-1)\,,\nonumber\\
K_1&=-\frac{323+2593\varpi+8128\varpi^2+12658\varpi^3+10235\varpi^4+4157\varpi^5+786\varpi^6}{4\zh^5(1+2\varpi)^3(2+3\varpi)}\nonumber\\
&=-\frac{72}{\zh^5}+\mathcal{O}(\varpi-1)\,.\label{ExK0K1}
\end{align}
and for $\zeta_{1,2}$ in (\ref{ZetaDefs}) we obtain
\begin{align}
\zeta_1&=-\frac{\zh^5(1+2\varpi)^3(2+\varpi-3\varpi^2)}{\ma(323+2593\varpi+8128\varpi^2+12658\varpi^3+10235\varpi^4+4157\varpi^5+786\varpi^6)}\nonumber\\
&=\frac{\zh^5}{288\ma}\,(\varpi-1)+\mathcal{O}((\varpi-1)^2)\,,\nonumber\\
\zeta_2&=\frac{(2 \varpi +1)^5 \zh^9}{\ma (3+4 \varpi) \left(786 \varpi ^6+4157
   \varpi ^5+10235 \varpi ^4+12658 \varpi ^3+8128 \varpi ^2+2593 \varpi +323\right)^3 }\nonumber\\
   &\hspace{0.5cm}\times\big(68694+901511 \varpi+5091799 \varpi ^2+15949627 \varpi ^3+29316755 \varpi ^4+28823217
   \varpi ^5\nonumber\\
   &\hspace{0.8cm}+4572149 \varpi ^6-24913951 \varpi ^7-32758909 \varpi ^8-19894452 \varpi ^9-6289848 \varpi ^{10}-866592 \varpi ^{11}\big) \nonumber\\
   &=-\frac{25\zh^9}{124416\ma}\,(\varpi-1)+\mathcal{O}((\varpi-1)^2)\,.\label{Exzeta}
\end{align}
Inserting (\ref{ExK0K1}) and (\ref{Exzeta}) into (\ref{K2solt}) we find
\begin{align}
K_2=\left\{\begin{array}{lcl}\frac{252}{\zh^6}+\frac{5116}{35\zh^6}\,(\varpi-1)+\mathcal{O}(\varpi-1) & \text{if} & u=+1\text{ and }\varpi\geq 1\,,\\[6pt] 
\frac{252}{\zh^6}+\frac{5116}{35\zh^6}\,(\varpi-1)+\mathcal{O}(\varpi-1) & \text{if} & u=-1\text{ and }\varpi\leq 1\,.\end{array}\right.\label{LimitAmbiguity}
\end{align}
Thus, depending on how the limit $\varpi\to 1$ is taken, \emph{i.e.} depending on the sign of $\xi_2$, only one choice of the sign $u$ in (\ref{K2solt}) yields a viable solution, which moreover leads to a finite limit for the Kretschmann scalar as a function of $\varpi$. Notice also, that in this case (\ref{ExK0K1}) and (\ref{LimitAmbiguity}) are compatible with the expansion of the Kretschmann scalar for the (classical) Schwarzschild space-time
\begin{align}
K_{\text{class}}=\frac{48\ma^2}{\co^6}=\frac{12}{\zh^4}-\frac{72}{\zh^5}\,(\co-\zh)+\frac{252}{\zh^6}\,(\co-\zh)^2+\mathcal{O}((z-\zh)^3)\,,
\end{align}
where we have used $\zh=2\ma$ for the classical Schwarzschild black hole. 

\printbibliography
\end{document}